\documentclass[prc,showpacs,showkeys,amssymb,amsmath,onecolumn,preprint]{revtex4}
\usepackage{amssymb}
\usepackage{amsmath}
\usepackage[pdftex]{graphicx}
\usepackage{epsfig}
\usepackage{epstopdf}
\usepackage{color}
\usepackage{slashed}
\newcommand{\be}{\begin{equation}}
\newcommand{\ee}{\end{equation}}
\newcommand{\ben}{\begin{eqnarray}}
\newcommand{\een}{\end{eqnarray}}
\newcommand{\lb}{\label}

\begin{document}
\title{Inverse magnetic catalysis and size-dependent effects on the chiral symmetry restoration}

\author{Luciano M. Abreu}
\email[]{luciano.abreu@ufba.br}
\affiliation{Instituto de F\'{\i}sica, Universidade Federal da Bahia, 40170-115, Salvador, BA, Brazil}

\author{Emerson B. S. Corr\^ea}
\email[]{emersoncbpf@gmail.com}
\affiliation{Faculdade de F\'isica, Universidade Federal do Sul e Sudeste do Par\'a, 68505-080, Marab\'a, PA, Brazil}

\author{Elenilson S. Nery}
\email[]{elenilsonnery@hotmail.com}
\affiliation{Instituto de F\'{\i}sica, Universidade Federal da Bahia, 40170-115, Salvador, BA, Brazil}

\begin{abstract}

We investigate the combined finite-size and thermo-magnetic effects on the properties of the quark matter, in the context of the two-flavored Nambu--Jona-Lasinio model. In particular, by using the mean-field approximation and the Schwinger proper time method in a toroidal topology with periodic or antiperiodic conditions, we evaluate the chiral phase transition, the constituent quark mass and the thermal and spatial susceptibilities under the change of the size, temperature and strength of external magnetic field. To take into account the inverse magnetic catalysis phenomenon, we make use of a recently proposed magnetized coupling constant. The findings suggest that the observables are strongly affected by the variation of the variables and also by the periodicity of the boundary conditions, with the final outcomes depending on the balance of these competing phenomena.

\end{abstract}
\keywords{Nambu--Jona-Lasinio model; inverse magnetic catalysis; finite-size effects}
\pacs{11.10.Wx, 12.39.-x, 12.38.Aw}

\maketitle
%

\section{Introduction}

Thanks to the theoretical and experimental advance made during the last decades, the phase diagram experienced by the strongly interacting matter is now better delineated. The observation of the deconfined state called quark-gluon plasma (QGP) in heavy-ion collisions~\cite{rev-qgp,Prino:2016cni,Pasechnik:2016wkt}, together with other breakthroughs, have been relevant steps towards its more compelling characterization. However, despite these progresses,  it persists as a hot research topic due to its rich and complex structure.

In particular, one of the largely investigated subjects is the dynamical chiral symmetry phase transition suffered by the system when submitted to extreme conditions, like high temperature and/or chemical potential (baryonic density). Besides, in a heavy-ion collision or in a compact star environment, other thermodynamic variables appears as relevant for the assessment of the chiral phase diagram, as the magnetic background~\cite{Kharzeev,Skokov:2009qp,Chernodub:2010qx,Ayala1,Tobias,Heber,MAO,Ayala2,Mamo:2015dea,Farias:2016gmy,Pagura,Magdy,Zhang:2016qrl,Ayala0,Wang:2017vtn,Martinez:2018snm,Mao:2018dqe,Avancini:2018svs,Avancini:2019wed,Abreu:2019czp,Ghosh:2021dlo,Abreu:2021btt}. In the case of colliders such as the RHIC and LHC, it is presumed that the magnetic field strength $\omega = e H$ has magnitude in the hadronic scale: $\omega = e H \sim 1-15\; m_{\pi}^2$ ($ m_{\pi} = 135-140$ MeV is the pion mass). Therefore, several predictions have been suggested in order to estimate the magnetic field influence. 
Some interesting physical effects have been proposed, such as the enhancement of the chiral condensate with the magnetic field - the magnetic catalysis (MC); as well as the restoration of chiral symmetry and suppression of the mentioned condensate - the inverse magnetic catalysis (IMC)~\cite{Tobias,MAO,Mamo:2015dea,Pagura,Magdy,Ayala0,Avancini:2018svs,Bali:2012zg,Bali:2011qj,Farias:2014eca,Ferreira:2014kpa,Farias:2016gmy,Martinez:2018snm,Ferreira:2017wtx,Ahmad:2016iez}. It should be remarked that lattice quantum chromodynamics (LQCD) calculations predict the MC effect at low temperatures, but yields the IMC close to the pseudocritical temperature. For a detailed discussion, we refer the reader to Ref.~\cite{Andersen:2021lnk}. 
 
At the same time, it has also been argued in literature that finite-volume effects might be taken into account on the phase structure of strongly interacting matter. This assumption relies on the idea that QGP-like systems produced in heavy-ion collisions are supposed to have a volume of the order of units or dozens of fm${}^3$~\cite{Bass:1998qm,Palhares:2009tf,Graef:2012sh,Shi:2018swj}.  
For example, in Ref.~\cite{Graef:2012sh} an analysis based on the Ultra relativistic Quantum Molecular Dynamics (UrQMD) transport approach has been done, showing that the volume of homogeneity before kinetic freeze-out (the lattest stage of a heavy-ion collision) for Au-Au collisions at center-of-mass energy $\sqrt{s} = 200$ GeV and for Pb-Pb collisions at $\sqrt{s} = 2.76$ TeV ranges between $25 \sim 250$ fm${}^3$ approximately. But according to Ref.~\cite{Palhares:2009tf} the volume of the smallest QGP system produced at RHIC (USA) could be  of the order of $(2 \, \mathrm{fm})^3$.
From this point of view, thermodynamic properties of strongly interacting matter might show dependence on finite-size effects depending on the range of volume considered and on the boundary conditions. Strictly speaking, in the bulk approximation the system can suffer a transition from chiral symmetry broken phase to the symmetric phase with the increase of the temperature and/or baryon chemical potential. But at a finite volume, the chiral symmetric phase is then enhanced~\cite{Shi:2018swj,Luecker:2009bs,Li:2017zny,Braun:2004yk,Braun:2005fj,Ferrer:1999gs,Abreu:2006,Ebert0,Abreu:2009zz,Abreu:2011rj,Bhattacharyya:2012rp,Bhattacharyya:2014uxa,Abreu6,Bhattacharyya2,Pan:2016ecs,Kohyama:2016fif,Damgaard:2008zs,Fraga,Abreu3,Ebert3,Abreu4,Magdy,Abreu5,Abreu7,Bao1,PhysRevC.96.055204,Samanta,Wu,Klein:2017shl,Shi,Wang:2018kgj,Abreu:2019czp,XiaYongHui:2019gci,Abreu:2019tnf,Das:2019crc,Zhao:2018nqa,Abreu:2021btt,Abreu:2020uxc}.
So, a natural question appears concerning the range where the bulk approximation remains valid for systems restricted to boundaries. 

In the end, according to the discussion above, one can ask about how the phase structure of a hot quark gas is influenced by the combined thermal-size-magnetic effects. 
We remark that in our previous work~\cite{Abreu:2021btt} this point has been studied in the context of the usual Nambu--Jona-Lasinio model~\cite{NJL,NJL1,Vogl,Klevansky,Hatsuda,Buballa}. The results indicate that the observables are affected by the conjoint effects of relevant variables. The inclusion of a magnetic background engendered the MC effect. However, as pointed out in other works these type of models are do not describe the IMC~\cite{Farias:2014eca,Ferreira:2014kpa}. The reason comes from the fact that in the NJL model the gluonic degrees of freedom, which play an inportant role in the suppression of the chiral condensate, are integrated out. So, the coupling of the the mentioned model does not behaves like the  the strong coupling  $\alpha_{s}$, which  decreases with the magnetic field strength and yields an effective weakening of the interaction between the quarks in the presence of an external magnetic field, and consequently the suppresion of the chiral condensate (i.e. the IMC effect). 

Hence, in the present work we intend to investigate the combination of IMC  with other effects in the scenario of NJL-like models, by performing some improvements in the approach present in ~\cite{Abreu:2021btt}. In particular, we will investigate the conjoint finite-size and thermo-magnetic effects on the properties of the quark matter, in the context of the two-flavor Nambu--Jona-Lasinio model. By using the mean-field approximation and the Schwinger proper time method in a toroidal topology with periodic (PBC) or antiperiodic (APBC)  boundary conditions, we analyze the gap equation solutions under the change of the size, temperature and strength of external magnetic field. The finite size effects are implemented according to the generalized Matsubara prescription. We employ a magnetic dependent coupling constant parametrized in consonance with the IMC predicted in LQCD, behaving in the same way of the strong coupling constant~\cite{Ferreira:2014kpa,Ferreira:2017wtx,Ahmad:2016iez}. 

We organize the paper as follows. In Section~II, we calculate the $(T,L,\mu,\omega)$-dependent gap equation from the NJL model in the mean-field approximation, using Schwinger's proper-time method and generalized Matsubara prescription. The results concerning the phase structure of the system, the critical coupling, the behavior of constituent quark mass and the spatial and chiral susceptibilities  are shown and analyzed in Section~III. Finally, Section~IV presents the concluding remarks.

\section{Formalism}

\subsection{The NJL model}

We start by introducing the density Lagrangian of the two-flavored NJL model, which is given by~\cite{NJL,NJL1,Vogl,Klevansky,Hatsuda,Buballa}
\begin{eqnarray}
	\mathcal{L}_{NJL} = \bar{q}(i{\slashed{\partial}}-\hat{m})q + G_{s} \left[\left(\bar{q}q\right)^{2}+\left(\bar{q} i\gamma_{5}\vec{\tau}q\right)^{2}\right],
	\label{L}
\end{eqnarray}
where $q = (u,d)^{T}$ represents the light quark field doublet $ (N_f = 2) $ with $ N_c = 3 $ colors, and $ \bar{q} $ its respective antiquark field doublet; $\hat{m} = {\rm diag}(m_{u},m_{d})$ is the current quark mass matrix; $G_{s} $ is the coupling constant of the scalar and pseudoscalar channel, and $\vec{\tau} $ are the Pauli matrices acting in isospin space. 
From now on it is assumed the isospin symmetry on the Lagrangian level, i.e. $m_u=m_d\equiv m$, and therefore $\hat{m}=m\mathbf{1}$.

%

Restricting the present analysis to the lowest-order evaluation of the phase structure, the mean-field (Hartree) approximation is employed. In view of this, the quark condensate $\phi\equiv\left\langle\bar{q}q\right\rangle$ is presumed to be the only allowed expectation value bilinear in the quark fields, engendering the linearization of the interaction terms in $\mathcal{L}_{\rm NJL}$  according to $(\bar{q}q)^2\approx2\phi(\bar{q}q)-\phi^2$. As result, neglecting the pseudoscalar contribution, it is possible to obtain from the Lagrangian density in Eq.~(\ref{L}) the expression
\begin{eqnarray}
	\mathcal{L}_{MF} = \bar{q}(i{\slashed{\partial}}-M)q - \frac{1}{4 G_{s}} (M-m)^2,
	\label{L1}
\end{eqnarray}
where $M$ denotes the constituent quark mass, given by
\be
M = m - 4 G_{s}\,\phi .
\label{masses}
\ee

To investigate the thermodynamic properties of the model we introduce the thermodynamic potential density at finite temperature $T$ and quark chemical potential $\mu$, which is defined by
\ben
\Omega (T, \mu) &  = & 
 - \frac{  1 }{\beta V } \ln{ {\rm Tr} \exp{ \left[ -\beta \int d^3 x \left( \mathcal{H} - \mu q^{\dagger } q \right)\right] } } , 
\lb{effpot1} 
\een 
where $\beta = 1/ T $, $\mathcal{H}$ is the Hamiltonian density (i.e. the Euclidean version of Lagrangian density $\mathcal{L}_{ MF}$) and ${\rm Tr}$ the functional trace over all states of the system (spin, isospin, color and momenta spaces). Therefore, the integration over fermion fields allows to write  the thermodynamic potential as
\be
\Omega(T,\mu) = \frac{1}{4 G_{s}} (M-m)^2 +  \Omega_{M}(T,\mu),  
\label{omega1}
\ee
where $\Omega_{M}(T,\mu_i)$ is the free Fermi-gas contribution,
\be
\Omega_{M}(T,\mu) = -\frac{N_c}{\beta}\sum_{n_{\tau}}\int\frac{d^3p}{(2\pi)^3}\mathrm{Tr}\ln\left[\slashed{p}\hat{1} -\mu\gamma^0-M \right].  
\label{omega2}
\ee
The sum $n_{\tau}$ stands for the sum over the fermionic Matsubara frequencies, $p_0 = i\omega_{n_{\tau}} = (2n_{\tau} + 1) \pi/\beta$. 

Thus, the gap equation can be derived by means of the minimization of the thermodynamic potential (\ref{omega1}) with respect to the constituent quark mass, and the physical solutions from the stationary points of the thermodynamic potential yield a useful expression for $\phi$, which after performing the trace over the internal spaces and appropriate manipulations can be given by 
\begin{eqnarray}
\phi & \equiv &  \left\langle  \bar{q}_i q_i \right\rangle = - 4  M  N_c \frac{1}{\beta} \sum _{n_{\tau}}  \int \frac{d^3 p}{(2 \pi)^3}\frac{1}{\tilde{\omega}_{n_{\tau}} ^{2} + \vec{p}^{2} + M^{2}},
\label{condensado}
\end{eqnarray}
where 
\be
 \tilde{\omega}_{n_{\tau}} = \frac{2\pi}{\beta }
	\left( n_{\tau} + \frac{1}{2} - i\frac{\mu \beta}{2\pi}\right). 
\ee 
In the next subsections we introduce the magnetic and finite-size effects.

%
%

\subsection{Generalized Matsubara prescription}

To include the finite-size effects on the phase structure of the model, we denote the Euclidean coordinate vectors by $x_E=(x_{\tau},x_1,x_2,x_3)$, where $x_{\tau}\in[0,\beta]$ and $x_j\in[0,L_j] \; (j=1,2,3)$ , with $L_j$ being the length of the compactified spatial dimensions.  Consequently, the Feynman rules in the sum-integral mixing in Eq.~(\ref{condensado}) should follow the so-called generalized Matsubara prescription~\cite{livro,PR2014,Emerson,Abreu:2019czp,Abreu:2021btt}, 
\begin{eqnarray}
\frac{1}{\beta }\sum_{ n_{\tau}=-\infty}^{\infty} \int\frac{d^3p}{(2\pi)^3}f(\tilde{\omega}_{n_{\tau}},\vec{p})\rightarrow \frac{1}{\beta L_1 L_2 L_3}\sum_{ n_{\tau} ,n_1,n_2,n_3=-\infty}^{\infty} f \left( \tilde{\omega}_{n_{\tau}}, \bar{\omega}_{n_1}, \bar{\omega}_{n_2},\bar{\omega}_{n_3} \right),\label{feynmanrule}
\end{eqnarray}
such that
\begin{eqnarray}
 {p}_{j}\rightarrow \bar{\omega} _{n_j} \equiv \frac{2\pi}{L_{j}}
	\left(n_{j}-b_{j}\right) \,,  \label{Matsubara}
\end{eqnarray}
where $n_{\tau}, n_{j} = 0,\pm 1 , \pm 2, \cdots$.  

Here it is worth noticing that due to the fermionic nature of the system, the Kubo-Martin-Schwinger conditions~\cite{livro} require anti-periodic boundary condition (ABC) in the imaginary-time coordinate ($b_{\tau} = -1/2$). With respect to the spatial compactified coordinates, however, there exists no fundamental constraints, with the choice of periodicity depending on the physical interest (see a detailed discussion in~\cite{Klein:2017shl,Abreu:2019czp,Abreu:2021btt}). Accordingly, the parameters $b_{j}$ in Eq.~(\ref{Matsubara}) can assume the values 0 or $-1/2$ for periodic boundary conditions (PBC) or anti-periodic boundary conditions (ABC), respectively. One fundamental consequence is concerning the spacetime permutation symmetry. The case of ABC in spatial compactified coordinates causes the physical equivalence of Euclidean space and time directions, keeping the permutation symmetry among them. As a result, the assumption of a temperature-independent vacuum coupling constant yields its spatial-independence as well. Contrarily, the periodic condition PBC breaks this permutation symmetry, and such spatial-dependence cannot be neglected in principle.

\subsection{Schwinger proper-time method}

The thermodynamic potential and the gap equations are treated using the Schwinger proper-time method~\cite{Schwinger,DeWitt1,DeWitt2,Ball,Abreu:2019czp,Abreu:2021btt}, in which the kernel of the propagator in Eq.~({\ref{condensado}}) is rewritten as 
\begin{eqnarray}
	\frac{1}{\tilde{\omega}_{n_{\tau}} ^{2}  +\vec{p}^{\,2}+M^{2}}=	\int_{0}^{\infty} d S \exp\left[- S \left( \tilde{\omega}_{n_{\tau}} ^{2} +\vec{p}^{\,2}+M^{2} \right) \right],
	\label{proptime}
\end{eqnarray}
where $S$ is the so-called proper time. Therefore, by employing  Eq.~({\ref{proptime}}) and the generalized Matsubara prescription~(\ref{Matsubara}) into ({\ref{condensado}}), and after some manipulations, the quark chiral condensate can be reexpressed as
\begin{eqnarray}
\phi(T,L_j,\mu) &=& \frac{4 M N_{c}}{\beta L_{1}L_{2}L_{3}}\,N_{f}\int_{0}^{\infty} dS \exp[-S(M^{2}-\mu^{2})]\,\theta_{2}\left[\frac{2\pi\mu S}{\beta}\,;\,\exp\left(-\frac{4\pi^2 S}{\beta^2}\right)\right] \nonumber \\
&&\times \prod_{j=1}^{3} \theta_{2}\left[0\,;\,\exp\left(-\frac{4\pi^2 S}{L_{j}^2}\right)\right]
\label{Ch_Cond_ABC}
\end{eqnarray}
for ABC in spatial coordinates, and 
\begin{eqnarray}
\phi(T,L_j,\mu) &=& \frac{4 M N_{c}}{\beta L_{1}L_{2}L_{3}} \, N_{f}\int_{0}^{\infty} dS \exp[-S(M^{2}-\mu^{2})]\,\theta_{2}\left[\frac{2\pi\mu S}{\beta}\,;\,\exp\left(-\frac{4\pi^2 S}{\beta^2}\right)\right] \nonumber \\
&&\times \prod_{j=1}^{3} \theta_{3}\left[0\,;\,\exp\left(-\frac{4\pi^2 S}{L_{j}^2}\right)\right]
\label{Ch_Cond_PBC}
\end{eqnarray}
for PBC; the $ \theta_{2} $ and $ \theta_{3} $ are the Jacobi theta functions, defined as~\cite{Bellman,Mumford}:
\begin{eqnarray}
\theta_{2}(u;q) & = &  2 \sum_{n=0}^{+\infty} q^{(n+1/2)^2}\cos[(2n+1)u], \\ \nonumber
\theta_{3}(u;q) & = & 1 + 2 \sum_{n=1}^{+\infty} q^{n^2}\cos(2nu).
\end{eqnarray}
%

%

We simplify the present study by fixing $L_{i}=L$. The bulk and zero temperature limits ($L_{j} \rightarrow \infty$ and  $\beta \rightarrow \infty$) are obtained by performing the inverse correspondence of the Matsubara prescription properly. 

In the next subsection the magnetic effects will be included.

%

\subsection{Inclusion of magnetic effects}

Now we consider the system under the influence of an external magnetic background. The  magnetic effects are implemented by minimal coupling prescription in Eq.~(\ref{L}), namely: $\partial_{\mu} \rightarrow \partial_{\mu} + \mathrm{i}\, e \, \hat{Q_{f}} A_{\mu}$, where $A_{\mu}$ is the four-potential and $\hat{Q_f}$ is the quark electric charge of flavour $f$, being $Q_u = - 2 Q_d = 2 /3$. The Landau gauge $A^{\mu } = (0,0,xH,0)$ is chosen, which gives a homogeneous and constant magnetic field $H$ along to $z$ direction. As a result, 
the constituent mass in Eq.~(\ref{masses}) is rewritten as  
\be
M = m - 2 {G_s} \sum_{f =u,d}  \phi_f (\omega),
\label{massaH}
\ee 
where the magnetic-dependent chiral condensate $\phi_f (\omega) $, after the Wick rotation in momenta space, is given by 
\ben
\phi_f (\omega) = 4 N_c M  \frac{|Q_f|\omega}{ 2 \pi } \sum_{\ell = 0}^{+\infty}\sum_{s=\pm1}^{}  \int \frac{dq_\tau}{(2\pi)} \frac{dq_z}{(2\pi)}\frac{ 1 }{q_{\tau}^{2}+q_{z}^{2} + |Q_f| \omega (2\ell+1 - s)+ M^{2}},
\een
with $\omega \equiv e H $ representing the cyclotron frequency, $ s=\pm 1$ the spin polarization and $\ell$ the Landau levels. 

In the following, we apply the recipe presented in the previous subsections, by employing the Matsubara generalized prescription~(\ref{Matsubara}) to account for finite temperature, chemical potential and size effects, and the Schwinger proper time parametrization. Then, after performing the sum over the spin polarizations $s$ and the geometrical series in $\ell$, we obtain 
\begin{eqnarray}
\phi_f(\omega,T,L_{z},\mu) &=& \frac{2N_{c}M  \omega }{\pi\beta L_{z}} \int_{0}^{\infty} dS \exp[-S(M^{2}-\mu^{2})]\,\theta_{2}\left[\frac{2\pi\mu S}{\beta}\,;\,\exp(-4\pi^2 S/\beta^2)\right] \nonumber \\
&&\times \theta_{2}\left[0\,;\,\exp(-4\pi^2 S/L_{z}^2)\right]\left[\frac{}{}|Q_f|\coth(|Q_f| \omega S)\right],
\label{phi3magABC}
\end{eqnarray}
for ABC in $z$ direction, and 
\begin{eqnarray}
	\phi_f(\omega,T,L_{z},\mu) &=& \frac{2N_{c}M  \omega }{\pi\beta L_{z}}\int_{0}^{\infty} dS \exp[-S(M^{2}-\mu^{2})]\,\theta_{2}\left[\frac{2\pi\mu S}{\beta}\,;\,\exp(-4\pi^2 S/\beta^2)\right] \nonumber \\
	&&\times \theta_{3}\left[0\,;\,\exp(-4\pi^2 S/L_{z}^2)\right]\left[\frac{}{}|Q_f|\coth(|Q_f| \omega S)\right],
	\label{phi3magPBC}
\end{eqnarray}
for PBC. 



\subsection{Magnetic-dependent coupling constant} 
As already discussed in literature, the presence of an external magnetic field engenders relevant physical effects, such as the magnetic catalysis (MC) and the inverse magnetic catalysis (IMC). The former is characterized by the enhancement of the chiral condensate with the magnetic field, whereas the latter is related to the suppression of the condensate, which by its turn yields the decrease of the
pseudocritical chiral transition temperature~\cite{Bali:2012zg,Bali:2011qj,Farias:2014eca,Ferreira:2014kpa,Farias:2016gmy,Martinez:2018snm}. Lattice quantum chromodynamics
(LQCD) calculations are in consonance with the MC effect at low temperatures, but predict the IMC close to the pseudocritical temperature. For a detailed discussion, we refer the reader to Ref.~\cite{Andersen:2021lnk}. 
 
In the context of NJL-like models, the inclusion of a magnetic background properly characterizes the MC effect, but is unsuccessful to describe the IMC. As explained in~\cite{Farias:2014eca,Ferreira:2014kpa},  in this framework the quarks assume pointlike effective interactions and the gluonic degrees of freedom integrated out. However, in the region of low momenta relevant for chiral symmetry breaking, the screening effect of the gluon
interactions plays a important role, since it suppresses the chiral condensate. The gluon acquires an effective mass proportional $m_G\propto\sqrt{N_f \alpha_s\omega}$ in this region, while the running strong coupling constant behaves with the magnetic field according to~\cite{Miransky},
\be
\alpha_{s}(\omega)=\frac{1}{\left(\frac{11N_c-2N_f}{6\pi}\right)\ln\left(\frac{\omega}{\Lambda_{QCD}^2}\right)},
\label{Coupling1}
\ee
Thus, the strong coupling  $\alpha_{s}$ decreases with the magnetic field strength, which yields an effective weakening of the interaction between
the quarks in the presence of an external magnetic field, and consequently the suppresion of the chiral condensate (i.e. the IMC effect). 
Then, keeping in mind that the coupling $G_{s}$ in NJL model should be seen as $\propto \alpha_s$, it must behave in the same way with external magnetic field. As a consequence, in order to reproduce the pseudocritical temperature for the chiral transitions obtained in LQCD calculations~\cite{Andersen:2021lnk}, we follow the ansatz reported in Ref.~\cite{Ferreira:2014kpa,Ferreira:2017wtx} and adopt a magnetic-dependent NJL coupling given by 
\be
G \left( \zeta \right) = \  G_{s} \ \left(\frac{1+a\,\zeta^{2}+b\,\zeta^{3}}{1+c\,\zeta^{2}+d\,\zeta^{4}}\right),
\label{Coupling2}
\ee
where $\zeta \equiv \omega/ \Lambda_{QCD}^2$, $\Lambda_{QCD} = 0.300$~GeV, $a = 0.0108805 , b = -1.0133 \cdot 10^{-4} , c = 0.02228 ,  d = 1.84558 \cdot 10^{-4}$. At zero magnetic field this construction coincides with the model described in previous subsections, i.e., $G(\omega=0)=G_s$. On the other hand at a very strong strong magnetic field we get a vanishing coupling constant. This behavior is therefore in consonance with the IMC effect.

It is important to note that there are other proposals for the coupling in literature trying to reproduce the IMC phenomenon~(see Ref.~\cite{Martinez:2018snm} for a discussion). In particular, there are distinct choices of the parametrization for the ansatz depicted in Eq.~(\ref{Coupling2}), which obviously depends on the model and  regularization employed. For example Refs.~\cite{Ferreira:2014kpa,Ferreira:2017wtx} work within the three-flavor NJL model with a sharp cutoff in three-momentum regularization scheme; on the other hand, Ref.~\cite{Ahmad:2016iez} makes use of the two-flavor NJL model with a modified confining proper-time regularization. Notwithstanding, our tests with different parametrizations showed that the choice above gives a reasonable description of the pseudocritical temperature for the chiral transition, and is in accordance with the purpose of this work concerning a first general attempt of analysis of conjoint finite-size and thermo-magnetic phenomena with IMC.

\subsection{Regularization procedure}

We must adopt some prescription to prevent the divergencies in the integrals over the proper time $S$. In our previous work~\cite{Abreu:2021btt}, where we have analyzed the combined finite-size and thermomagnetic effects on the properties of neutral mesons in a hot medium without IMC, the regularization of the proper time method has been done through the use of an ultraviolet cutoff $\Lambda$, according to the anzatz
\begin{eqnarray}
\int_{0}^{\infty}f(S)\,dS \rightarrow \int_{1/\Lambda^2}^{\infty}f(S)\,dS.
\label{cuttof}	
\end{eqnarray}

However, we would like to mention an extensive study performed in Ref.~\cite{Avancini:2019wed} concerning the regularization dependence in NJL-type models in the presence of intense magnetic and at zero temperature. The average and difference of the quark condensates using different regularizations have been calculated and compared with recent lattice results. Among them, the proper time method has been investigated in two situations: taking the so-called magnetic field independent regularization  (MFIR) procedure, where the finite magnetic contribution is disentangled from the non-magnetic infinite one and only the latter is regularized; as well as in the non-MFIR (nMFIR), in which the 
the magnetic and nonmagnetic vacuum contributions are entangled, as done in~\cite{Abreu:2021btt}. In the latter case (nMFIR), the integrals in the condensates are treated by employing Eq.~(\ref{cuttof}) above. In the MFIR, the condensates assume the form
\begin{eqnarray}
\phi_f(\omega,T,L_{z},\mu) &=& \frac{2N_{c}M  }{\pi} \frac{\pi}{4\pi^2} \left\{  \int_{1/\Lambda^{2}}^{\infty} \frac{dS}{S^{2}} \exp\left(-S \,M^{2}\right) \right. \nonumber \\
&+& \omega_{f}\left[ 2 \, \zeta^{\prime}(0,x_{f})+(1-2x_{f})\ln{x_{f}}+2x_{f}  \right] \nonumber \\
&+& \omega_{f} \sum_{\ell = 0}^{+\infty}\sum_{s=\pm1}^{} \left[\left[2^{2} \sum_{n_{z} = 1}^{+\infty} (-1)^{n_{z}} K_{0}\left(L_{z} n_{z}\sqrt{M^{2}+\omega_{f}(2\ell+1-s)}\right) \right] \right. \nonumber \\
&+&\left[2^{2} \sum_{n_{\tau} = 1}^{+\infty} (-1)^{n_{\tau}} \cosh{(n_{\tau}\beta\mu)}K_{0}\left(\beta \,n_{\tau}\sqrt{M^{2}+\omega_{f}(2\ell+1-s)}\right) \right] \nonumber \\
&+&\left[2^{3} \sum_{n_{\tau},n_{z} = 1}^{+\infty} (-1)^{n_{\tau}+n_{z}} \cosh{(n_{\tau}\beta\mu)} \right.\nonumber \\
&\times & \left.\left.\left.K_{0}\left(\sqrt{\left(\beta^2 n_{\tau}^2 + L_{z}^{2} n_{z}^{2}\right)\left(M^{2}+\omega_{f}(2\ell+1-s)\right)}\right)\right]\right]\right\},
\label{regabc}
\end{eqnarray}
for ABC, and
\begin{eqnarray}
\phi_f(\omega,T,L_{z},\mu) &=& \frac{2N_{c}M  }{\pi} \frac{\pi}{4\pi^2} \left\{  \int_{1/\Lambda^{2}}^{\infty} \frac{dS}{S^{2}} \exp\left(-S \,M^{2}\right) \right. \nonumber \\
&+& \omega_{f}\left[2 \, \zeta^{\prime}(0,x_{f})+(1-2x_{f})\ln{x_{f}}+2x_{f} \frac{}{}\right] \nonumber \\
&+& \omega_{f} \sum_{\ell = 0}^{+\infty}\sum_{s=\pm1}^{} \left[\left[2^{2} \sum_{n_{z} = 1}^{+\infty}  K_{0}\left(L_{z} n_{z}\sqrt{M^{2}+\omega_{f}(2\ell+1-s)}\right) \right] \right. \nonumber \\
&+&\left[2^{2} \sum_{n_{\tau} = 1}^{+\infty} (-1)^{n_{\tau}} \cosh{(n_{\tau}\beta\mu)}K_{0}\left(\beta \,n_{\tau}\sqrt{M^{2}+\omega_{f}(2\ell+1-s)}\right) \right] \nonumber \\
&+&\left[2^{3} \sum_{n_{\tau},n_{z} = 1}^{+\infty} (-1)^{n_{\tau}} \cosh{(n_{\tau}\beta\mu)} \right.\nonumber \\
&\times&\left.\left.\left.K_{0}\left(\sqrt{\left(\beta^2 n_{\tau}^2 + L_{z}^{2} n_{z}^{2}\right)\left(M^{2}+\omega_{f}(2\ell+1-s)\right)}\right)\right]\right]\right\},
\label{regpbc}
\end{eqnarray}
for PBC, where $x_{f} \equiv M^{2} / 2\omega_{f}$ and $\zeta^{\prime}(0,x_{f})$ is the derivative of the Hurwitz zeta function with respect to the first argument. The derivation of these expressions is presented in Appendix. 

Although Ref.~\cite{Avancini:2019wed} concludes that  4D-cutoff and the Pauli-Villars in the MFIR scheme are the best regularizations, in the case of the proper time method the nMFIR scheme presents a correct qualitative behavior and does not deviate too much from lattice results. Therefore, keeping in mind that the main aim of this work to perform a first general analysis of the combined finite-size and thermo-magnetic phenomena with IMC, for completeness we follow our previous work~\cite{Abreu:2021btt} and use an ultraviolet cutoff $\Lambda$ according Eq.~(\ref{cuttof}). We postpone a  detailed investigation about the dependence of the different regularization procedures for a further work, where we plan to investigate the properties of mesons in a hot medium with IMC.

\section{Results}

We devote this section to the analysis of the phase structure of the system, focusing on how it behaves with the change of the thermodynamic variables, especially on the behavior of constituent quark mass $ M $, obtained from the solution of the gap equation in Eq. (\ref{massaH}), and the chiral and spatial susceptibilities. Noticing that the present approach is intended to be
applied in a heavy-ion collision environment, characterized by a very low chemical potential $ \mu $, as a consequence
we concentrate here on the influence of the combined finite-size and thermo-magnetic effects.
The model introduced above carries the following free parameters: the coupling constant
$ G_s $, the ultraviolet cutoff $ \Lambda $, and the current quark mass $ m $. They are fixed in order to reproduce the observed hadron quantities at vacuum values of thermodynamic characteristics:  $T, 1/L$ and $\omega$.  (see Refs.~\cite{Klevansky,Kohyama:2016fif,Abreu:2021btt} for a detailed discussion). In this sense, we use the set of
parameters defined in Ref.~\cite{Abreu:2021btt} from the fitting of the pion mass and pion decay constant; explicitly: $ G_s = 5.691~ \mathrm{GeV}^{-2}$, $ \Lambda = 0.688~ \mathrm{GeV}$ and $m = 11.7~ \mathrm{MeV}$~\footnote{As discussed in Ref.~\cite{Abreu:2021btt}, the value fixed for $m$ is higher than those usually utilized in different parametrizations, due to the fact that most of authors prefer set the current quark mass close to the value estimated by quark models (see the PDG's review). Notwithstanding, several studies also prefer to fix the value of the constituent quark mass $M$, see for example~\cite{Cloet:2014rja,Zhang2016MPLA,Hutauruk:2018zfk}, yielding higher values for $ m $. Ref.~\cite{Kohyama:2016fif} summarizes other parametrizations with higher values of $m$ in the vacuum.
}.

\subsection{The critical coupling}

The starting point is the investigation of how the critical coupling $ G_s ^{(c)} $ establishing the regions of restoration or breaking of the chiral symmetry behaves in thermodynamic variable space. According to Refs.~\cite{Martinez:2018snm,Abreu:2021btt}, the value of $G_s ^{(c)}  $ where the trivial and  nontrivial  solutions ($M =0$ and $M\neq 0$)  bifurcate from  one  another can be determined by taking the derivative of the  gap  equation with respect  to $M$ at $M=0$. Then, the application of this method to Eq.~(\ref{masses}) in the bulk vacuum limit gives $G_s ^{(c)} = 2 \pi^2 / (3 N_f \Lambda^2) $.  
To include the thermo-magnetic and finite-size effects, one should extend this prescription to the modified gap equation (\ref{massaH}), which yields the following condition  for  criticality, 
\ben
1 = 2  {G_s ^{(c)} } \sum_{f =u,d} \left. \frac{\partial }{\partial M}  \left[\phi_f(\omega,T,L,\mu)  \right] \right|_{M=0}. 
\label{critcond}
\een 
As in Refs.~\cite{Martinez:2018snm,Abreu:2021btt}, we assume that the coupling is dressed by the thermo-magnetic medium with boundaries, and define the (pseudo-)critical dressed coupling $G_s ^{(c)} (L, T, \omega) $  as the value of $G_s $ needed to break chiral symmetry. In the bulk vacuum, the solution of Eq.~(\ref{critcond}) is obtained at $G_s ^{(c)} ( T, 1/L, \omega \rightarrow 0 ) \approx 3.46 \, \mathrm{GeV}^{-2}$, taking the cutoff introduced above. But in the regime of high temperatures experienced by the system in a heavy-ion collision environment, the values for $G_s ^{(c)} ( T ; 1/L, \omega \rightarrow 0 ) $ are most likely larger than that chosen in our parametrization. Therefore it is probably subcritical, in consonance with the context of heavy-ion collisions.

\begin{figure}
\centering
\includegraphics[width=0.45\columnwidth]{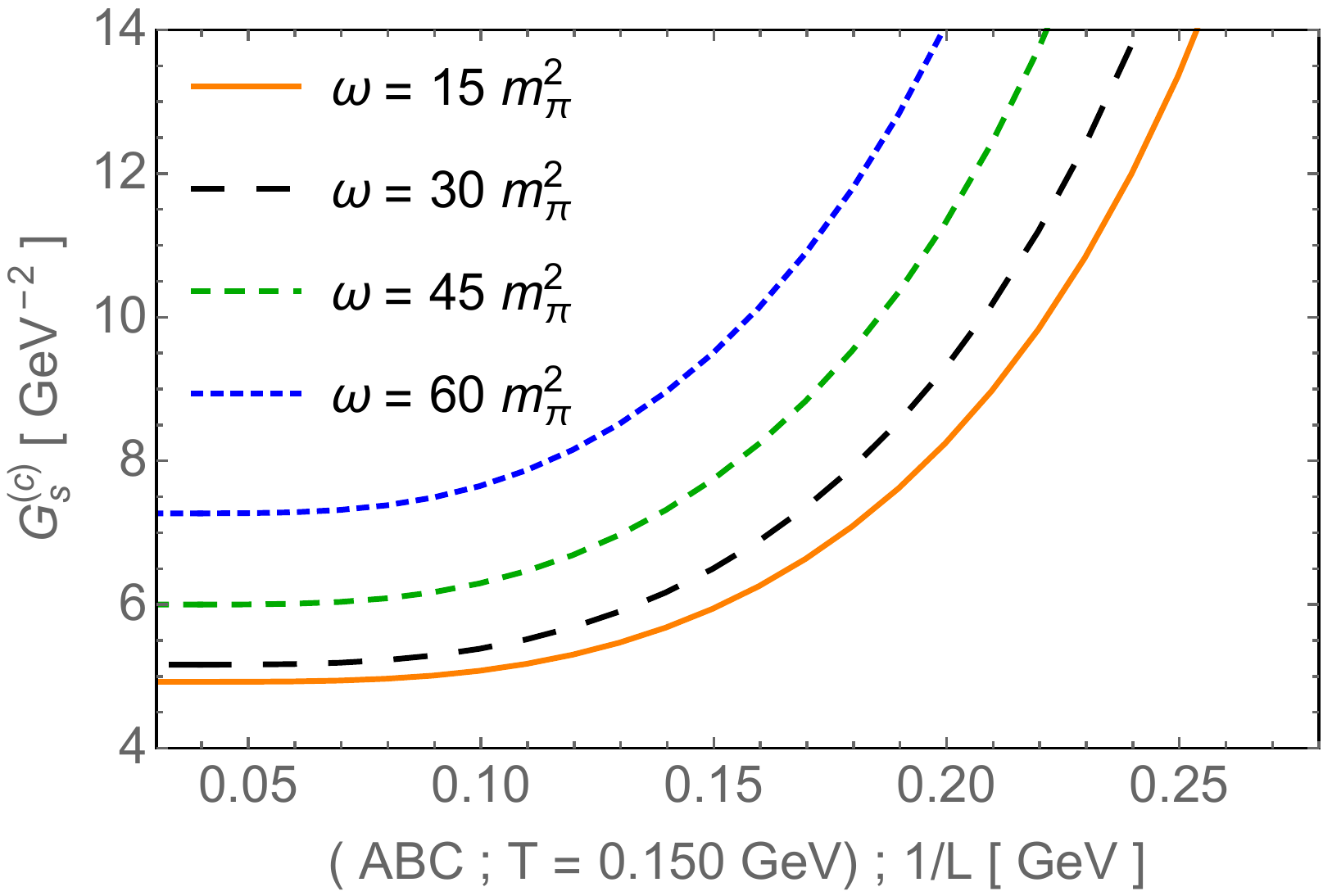}
\caption{Critical curves obtained from Eq.~(\ref{critcond}) as  a  function  of  the inverse of length $ 1/L $  in ABC  case, taking different values of  and cyclotron frequency $ \omega $ and at a given temperature $T$. Here we use the squared pion mass $ m_{\pi}^2 (\approx 0.018 \mathrm{GeV}^2)$ as scale for $ \omega $. }
\label{CriticalCouplingAPBC}
\end{figure}

\begin{figure}
\centering
\includegraphics[width=0.45\columnwidth]{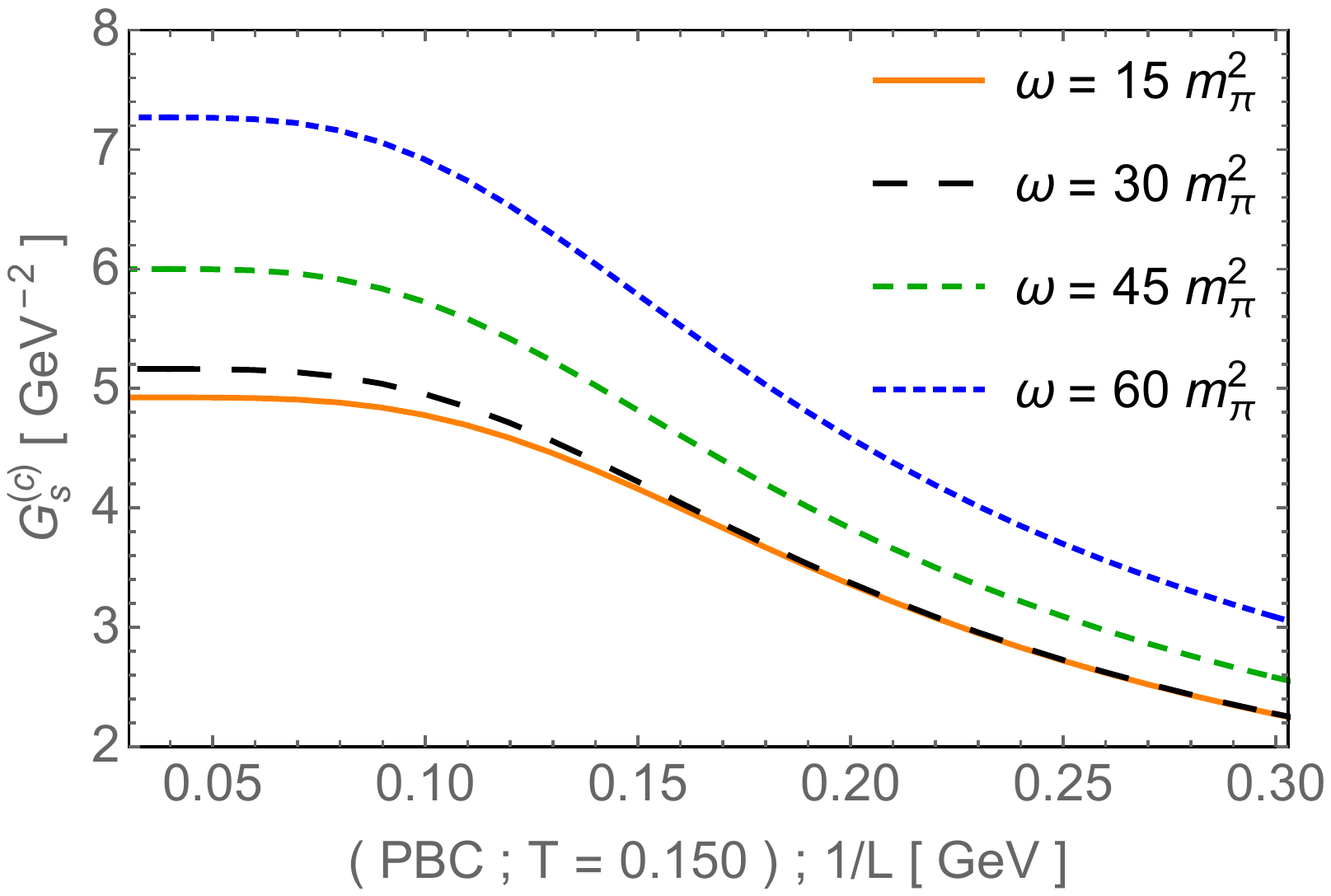}
\caption{The same as in Fig.~\ref{CriticalCouplingAPBC}, but in PBC case.  }
\label{CriticalCouplingPBC}
\end{figure}

In  Figs.~\ref{CriticalCouplingAPBC} and \ref{CriticalCouplingPBC} are plotted the critical curves obtained from Eq.~(\ref{critcond}) as  a  function  of  the inverse of length $ 1/L $,  taking different values of  and cyclotron frequency $ \omega $ and at a given temperature $T$, in both APBC and PBC cases. The domains above the curves correspond to the chirally broken region, where $G_s$ yields nonvanishing dynamical quark mass. We notice that the dependence of $G_s ^{(c)}$ with $\omega$  is different to that reported in~\cite{Abreu:2021btt}: both thermal and magnetic effects suppress the chiral broken phase. In particular, as the  magnetic field strength increases, $G_s ^{(c)} $ becomes bigger and the  critical temperature  moves  toward  smaller  values, manifesting the IMC effect. Concerning the periodicity of the boundaries, the parallelism between $1/L $ and $T$ in the ABC case engenders critical curves depending on $1/L$ analogously to $T$; the critical coupling will diverge at given values of $T$ and $1/L$, above which there is no chiral symmetry breaking. As a result, thermal and ABC size effects act as the magnetic ones. Moving on to the PBC case, the finding is different: $G_s ^{(c)} $ diminishes as $L$ decreases, causing the 
enhancement of the chiral broken phase. In the end, the critical coupling  presents a dependence on the boundary conditions. We explore the consequences more accurately in the next subsections.

\subsection{Constituent quark mass}

\begin{figure}
\centering
\includegraphics[width=0.45\columnwidth]{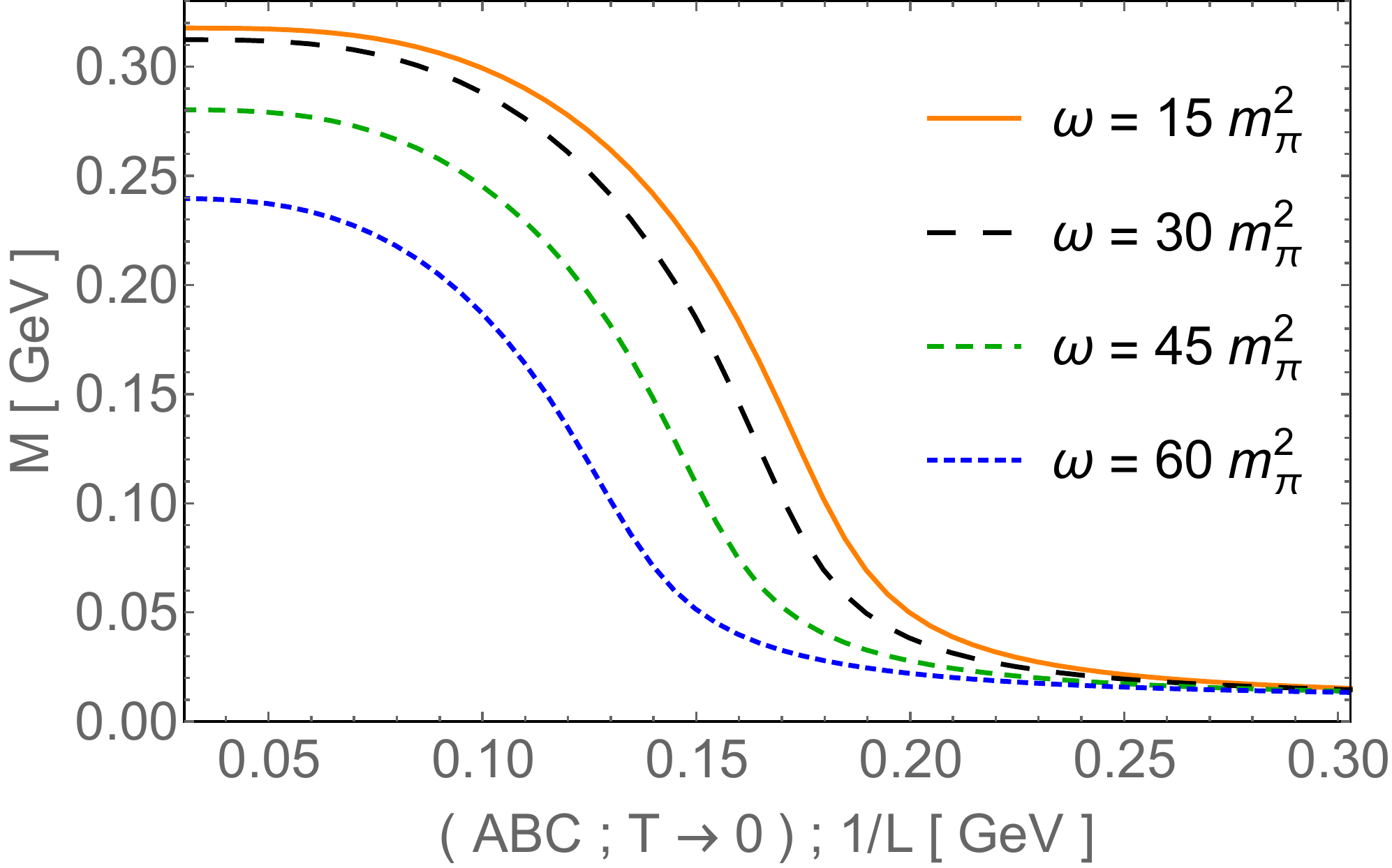}
\includegraphics[width=0.45\columnwidth]{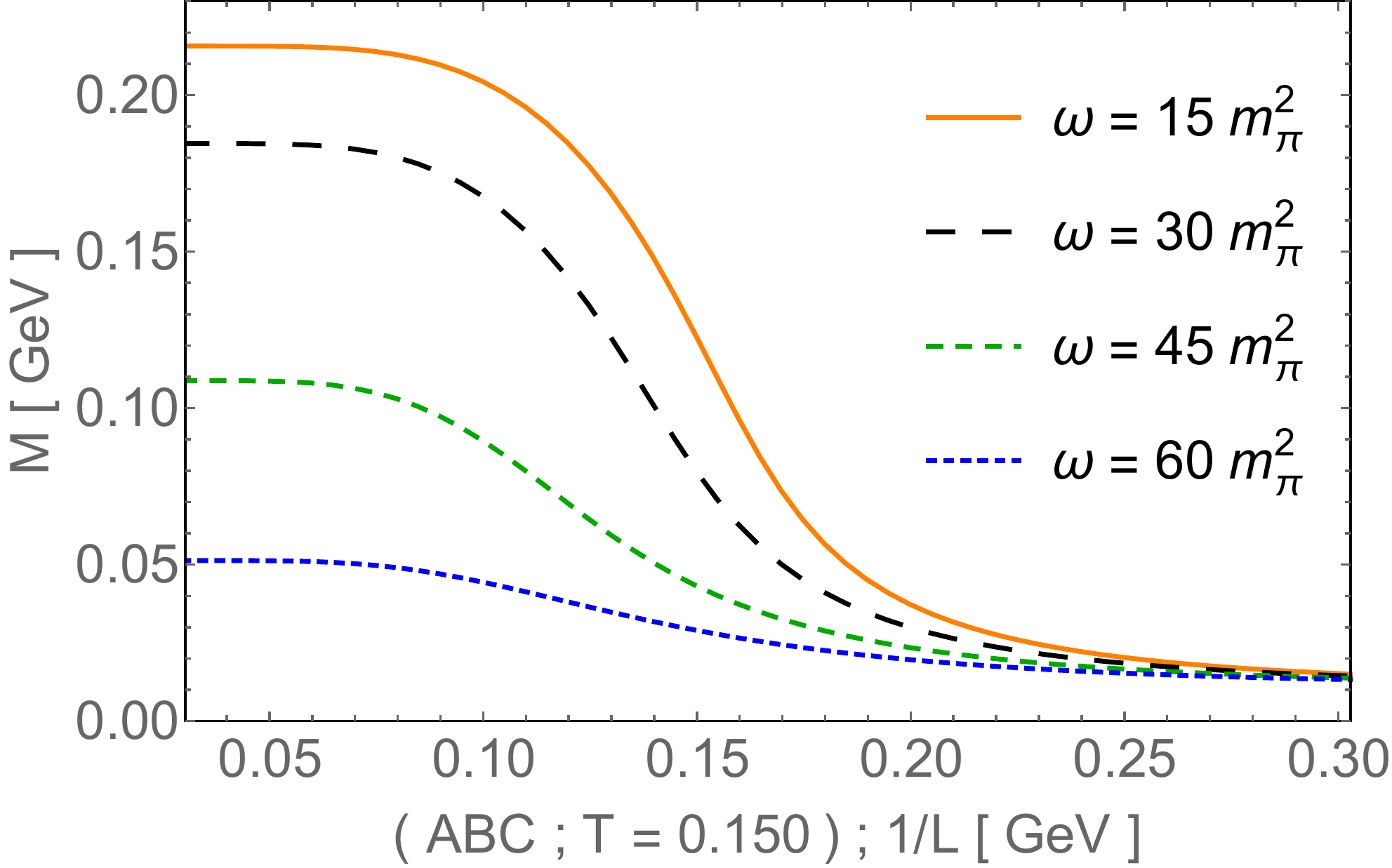}
\caption{Constituent quark mass $M$, obtained from the solutions of the gap equation in Eq.~(\ref{masses}), as a  function  of the inverse of length $ 1/L $  in ABC case, taking different values of  and cyclotron frequency $ \omega $ and temperature $T$. Here we use the squared pion mass $ m_{\pi}^2 (\approx 0.018 \mathrm{GeV}^2)$ as scale for $ \omega $. }
\label{MassaABC}
\end{figure}

\begin{figure}
\centering
\includegraphics[width=0.45\columnwidth]{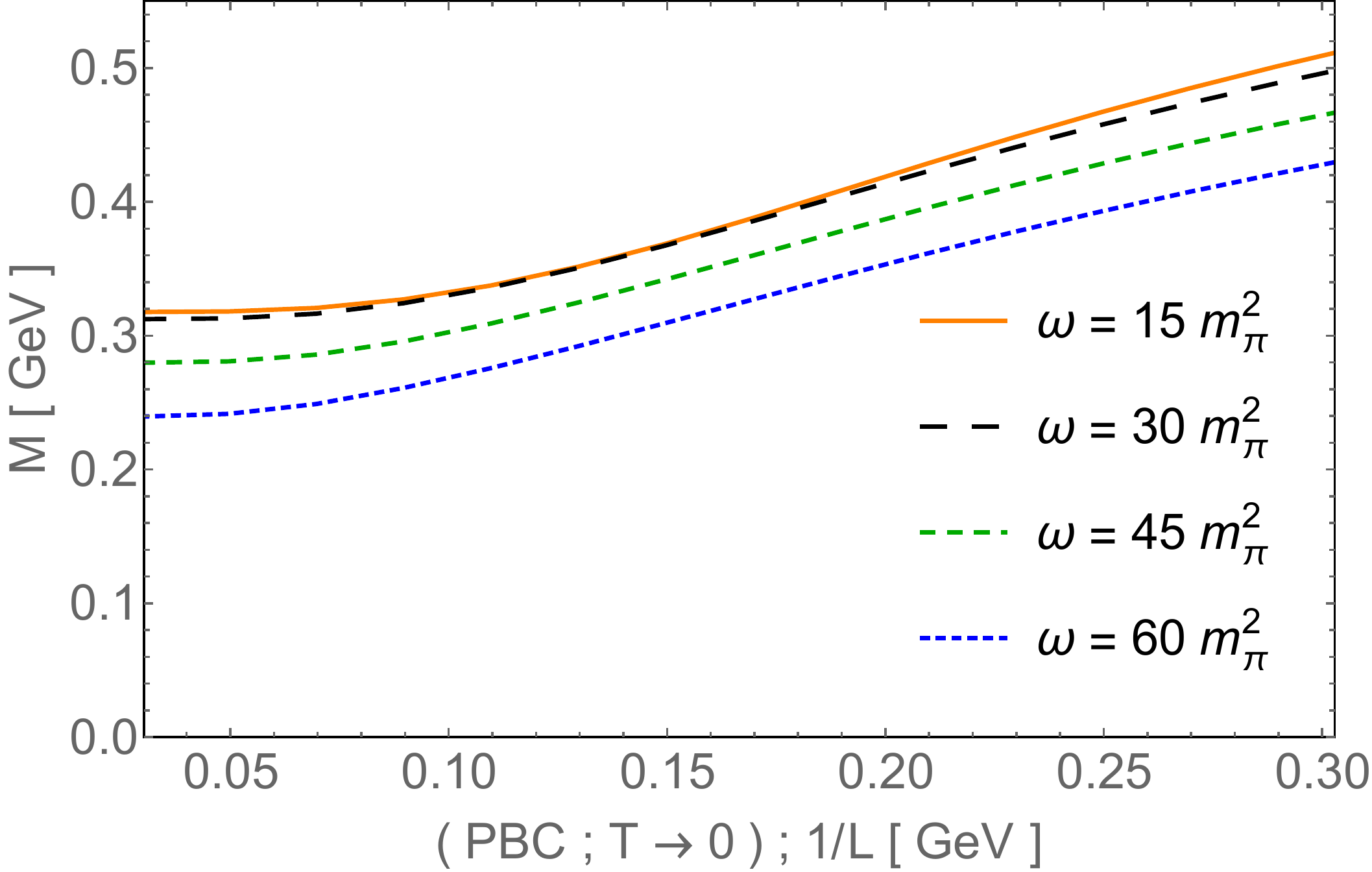}
\includegraphics[width=0.45\columnwidth]{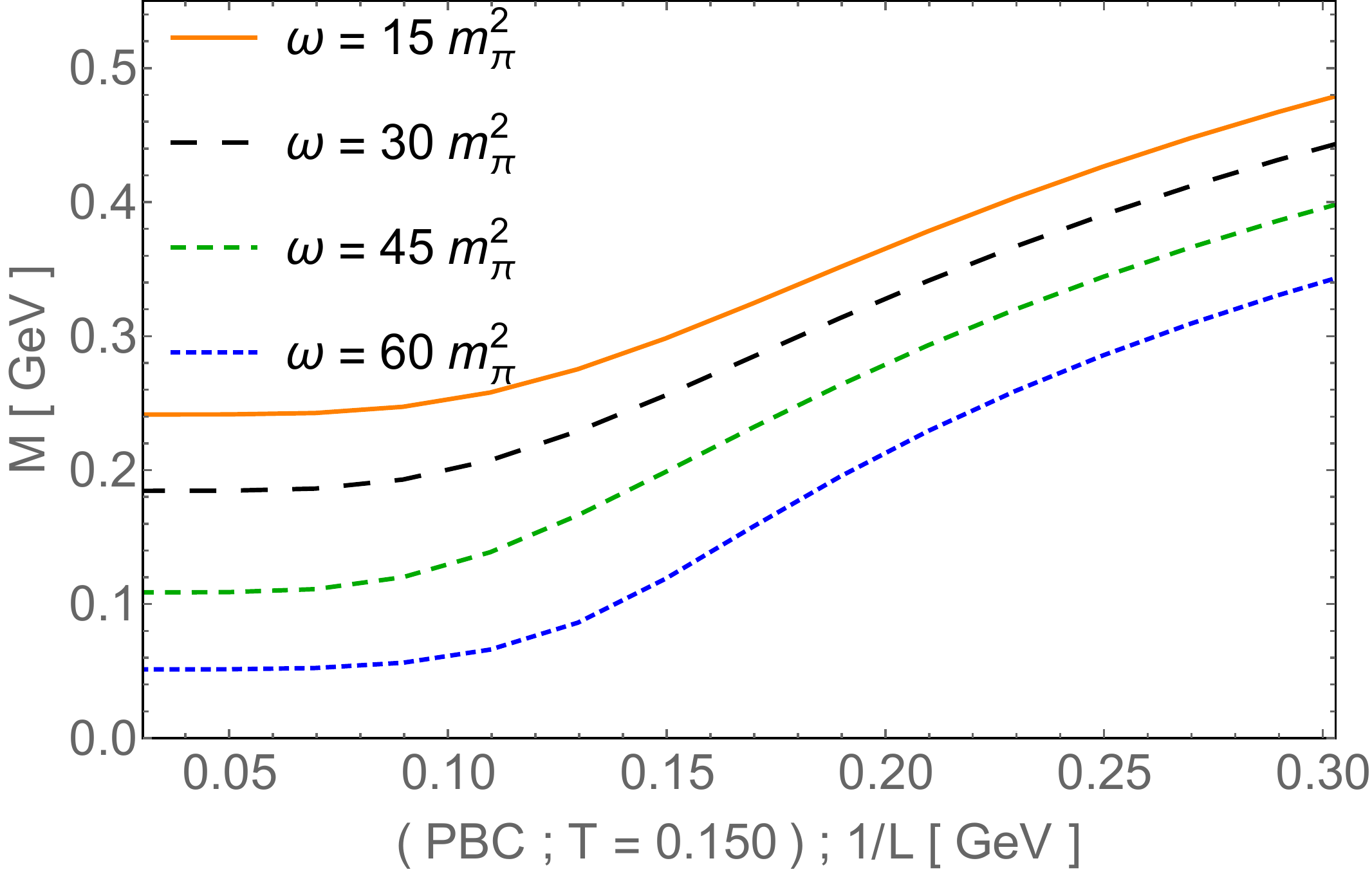}
\caption{The same as in Fig.~\ref{MassaABC}, but in PBC case. }
\label{MassaPBC}
\end{figure}

Here is analyzed the constituent quark mass $M$ obtained from the solutions of the gap equation in Eq.~(\ref{masses}) under the combined effects of boundaries, finite temperature and a magnetic background in the context described above. To this end, in Figs.~\ref{MassaABC} and \ref{MassaPBC} are plotted the values of $ M $ that are solutions of the gap equation in Eq.~(\ref{masses}) as a function of the different variables, using the coupling constant given by Eq.~(\ref{Coupling2}) and spatial boundary in ABC and PBC cases. It can be observed that in the region of lower values of $T, 1/L$ and $\omega$ there is no sizeable modifications on $M$, at which the vacuum mean-field approach holds as a good approximation. But the increase of any of these variables causes a huge fluctuation on $ M $. The dependence on the temperature appears as expected: at higher values of $ T $ the constituent mass falls smoothly to the current quark mass,  characterizing a crossover-like phase transition. In particular, at certain values of parameters the dressed mass converges to the current quark mass. The thermal effect has already been well investigated in literature, so we focus on the other variables. In the plots the IMC effect can be seen  in its ``pure" state: the growth of the magnetic field strength decreases $ M $ as well as the pseudo-critical temperature of the phase transition. 

Now we discuss the conjoint magnetic and boundary effects, starting with the ABC situation (Fig.~\ref{MassaABC}). In the range of magnetic field strength considered, the constituent quark mass lowers with the decreasing of the size, with the broken phase being inhibited and a crossover transition occurring. The typical range of $ L $ where this effect takes place is of the order of a few units of fm. So, ABC boundaries act similarly to the thermal effects in the phase diagram, because of the analogous ABC nature of $1/L $ and $1 / \beta  = T $.  Therefore, thermo-size-magnetic effects in the ABC scenario with $G_s(\zeta)$ disfavor the maintenance of long-range correlations, constrain and weaken the broken phase.

In PBC case (Fig.~\ref{MassaPBC}), however, the constituent quark masses acquire greater values with the augmentation of $1/L $, causing a reverse effect compared to temperature. It can be understood as folloes: from the generalized Matsubara prescription (\ref{Matsubara}), which states that the fermion fields with ABC must obey $( {p}_{j}\rightarrow \bar{\omega} _{n_j}  \geq \pi / L_{j})$, with ${p}_{j}$ being larger for smaller values of $L_j$. Noticing that the infrared contributions play an important role in the chiral symmetry breaking,  then in the chiral limit the chiral condensates defined in Eq.~(\ref{phi3magABC}) and~(\ref{phi3magPBC})  becomes zero at a sufficiently small size, generation the restoration of the chiral symmetry. But the PBC case does not have the restriction above mentioned for ${p}_{j}$; as a consequence the decrease of the size doe not give restoration of the symmetry. The correlation between the quarks is then favored for smaller size and provides a higher value of $\phi _f$ (see~\cite{Ishikawa:1996jb,Abreu:2020uxc,Abreu:2021btt} for more details). Thus, in this PBC context with $G_s(\zeta)$ the thermo-magnetic effects of restraining the broken phase compete with the finite-size effects of inducing its stimulation.

This sharp dependence of conjunction of finite-size and magnetic effects on the boundary conditions can be put in a more general perspective. In the framework of effective models, the ABC in spatial directions for the quark fields is the usual choice~\cite{Abreu:2020uxc}. Conversely, the PBC appears often in lattice QCD simulations in order to minimize empirically the finite-volume effects~\cite{Klein:2017shl,Magdy}.



\subsection{Thermal and spatial susceptibilities}

\begin{figure}
\centering
\includegraphics[width=0.45\columnwidth]{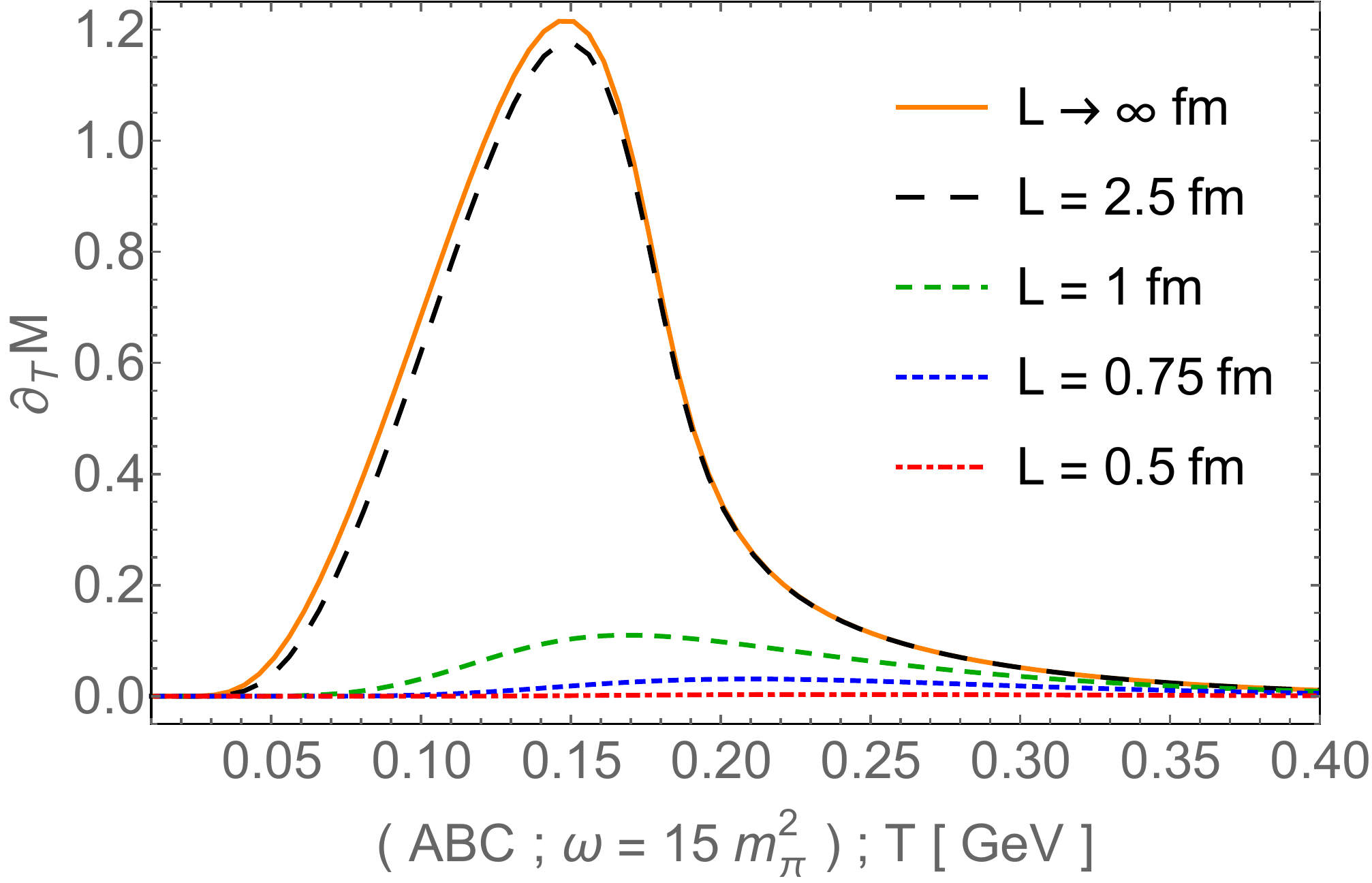}
\includegraphics[width=0.45\columnwidth]{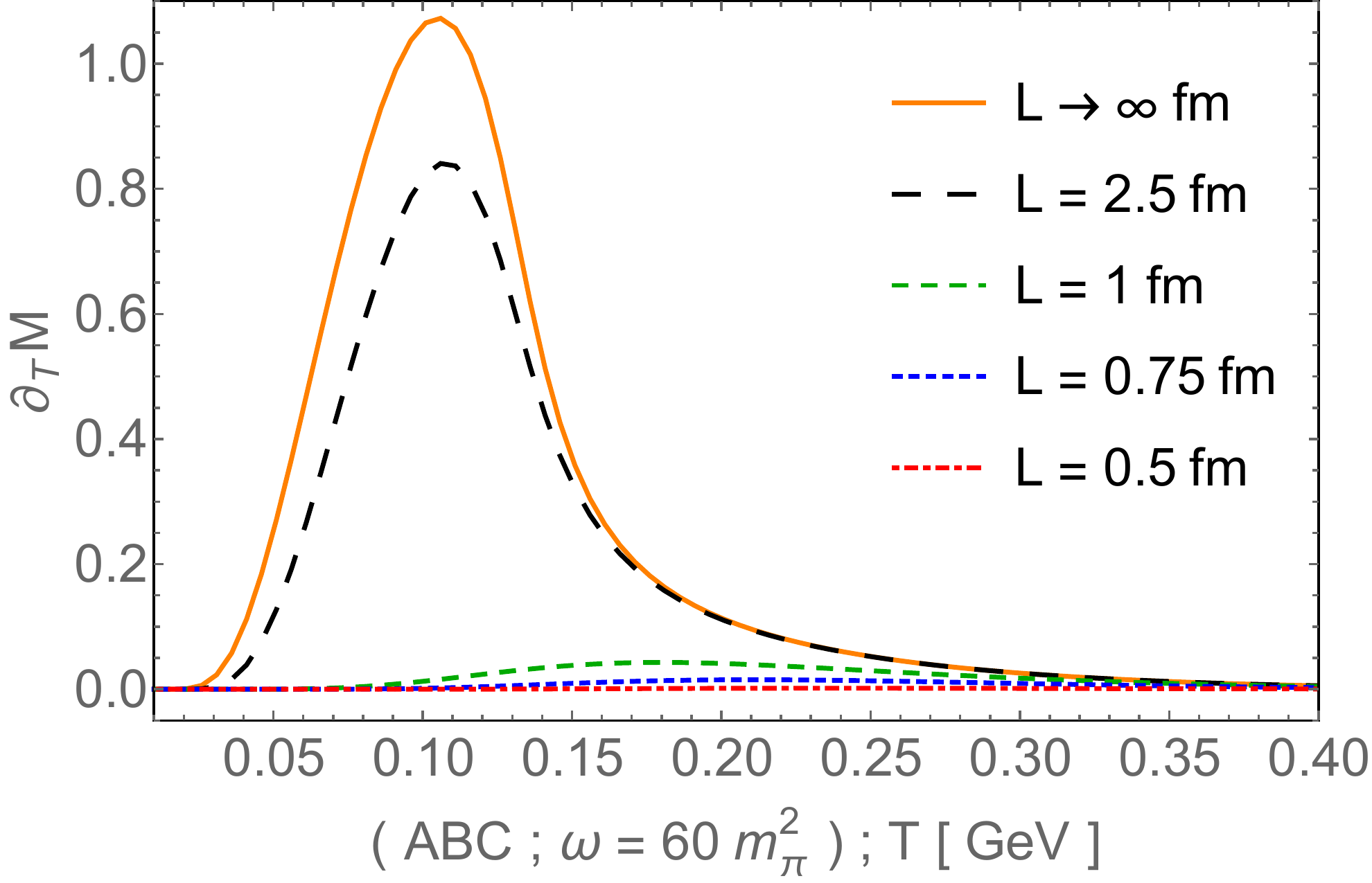}
\caption{Thermal susceptibility ($\partial M / \partial T$ ) as a function of the temperature $T$ in ABC case, taking different values of the cyclotron frequency $ \omega $ and length $L$. Here we use the squared pion mass $ m_{\pi}^2 (\approx 0.018 \mathrm{GeV}^2)$ as scale for $ \omega $. }
\label{Susc1}
\end{figure}

\begin{figure}
\centering
\includegraphics[width=0.45\columnwidth]{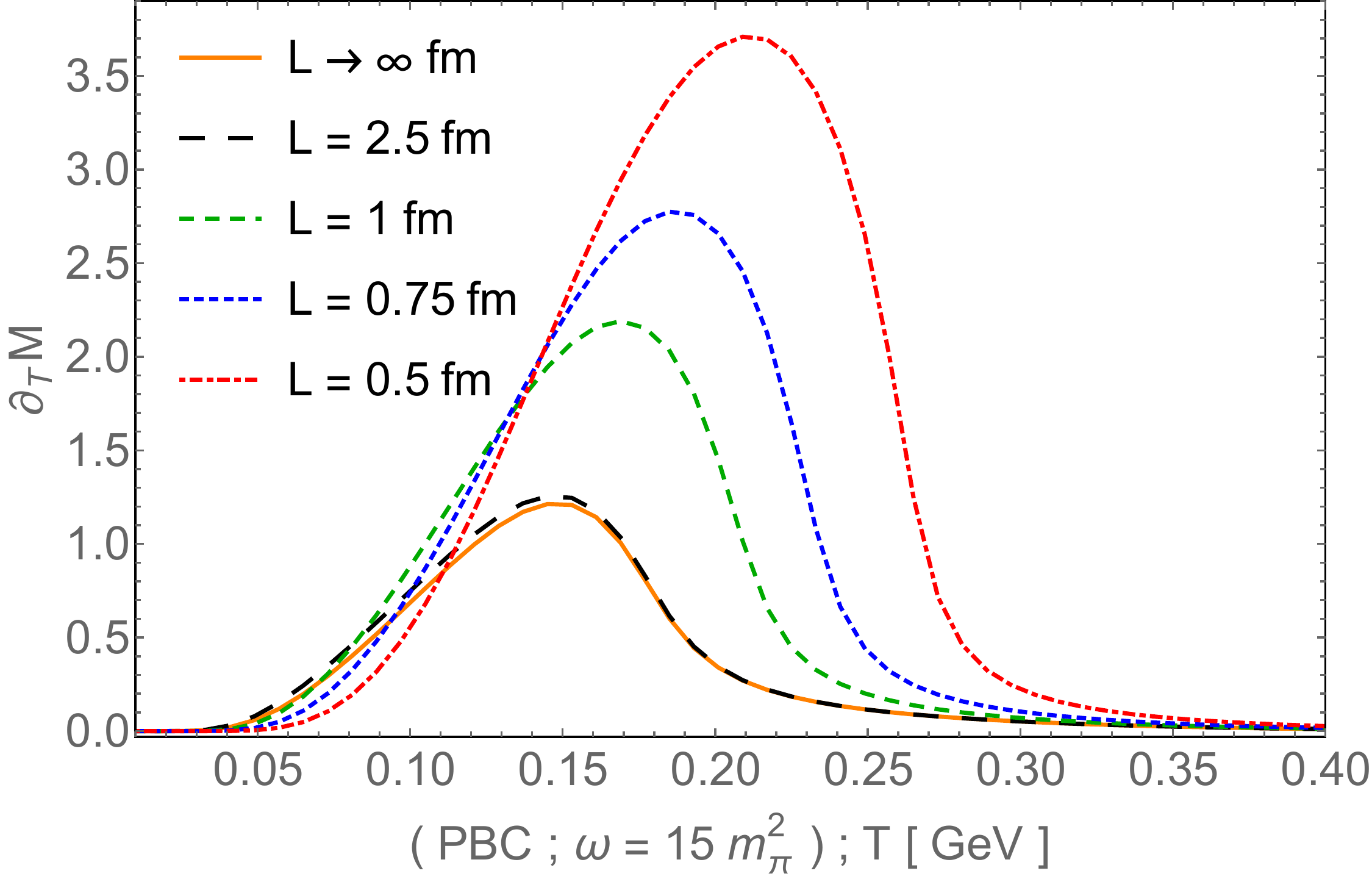}
\includegraphics[width=0.45\columnwidth]{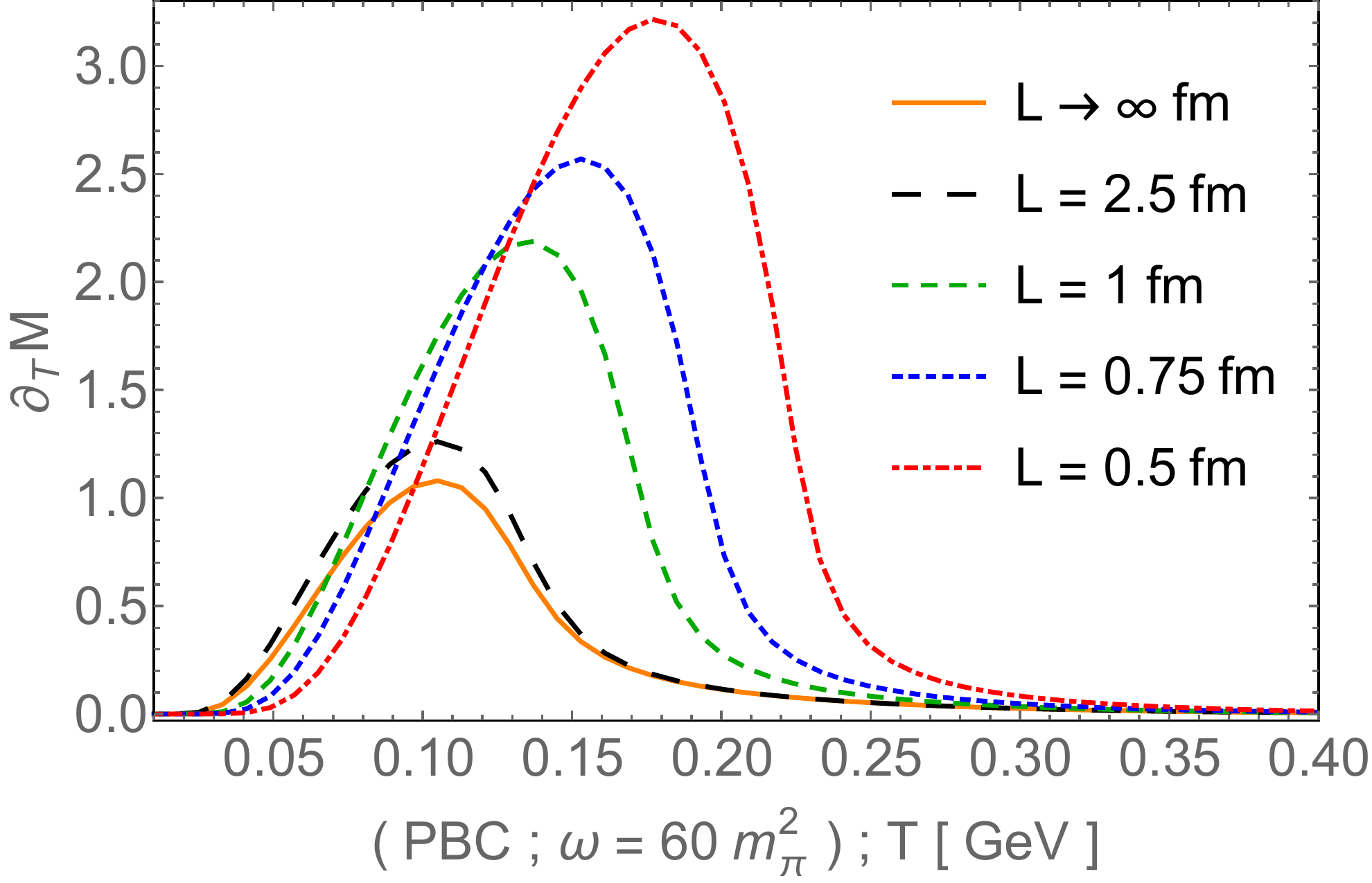}
\caption{The same as in Fig.~\ref{Susc1}, but in PBC case. }
\label{Susc2}
\end{figure}

\begin{figure}
\centering
\includegraphics[width=0.45\columnwidth]{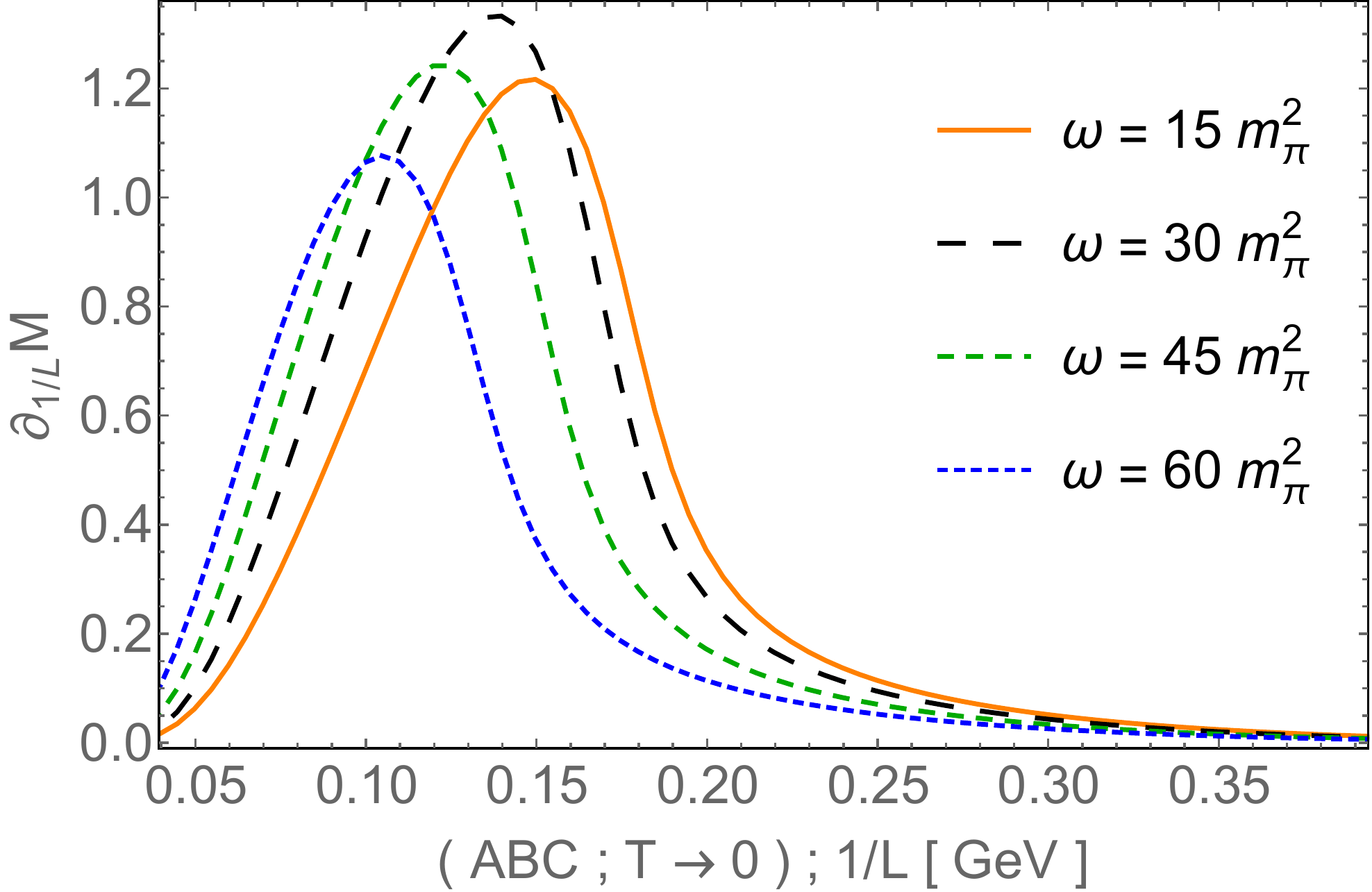}
\includegraphics[width=0.45\columnwidth]{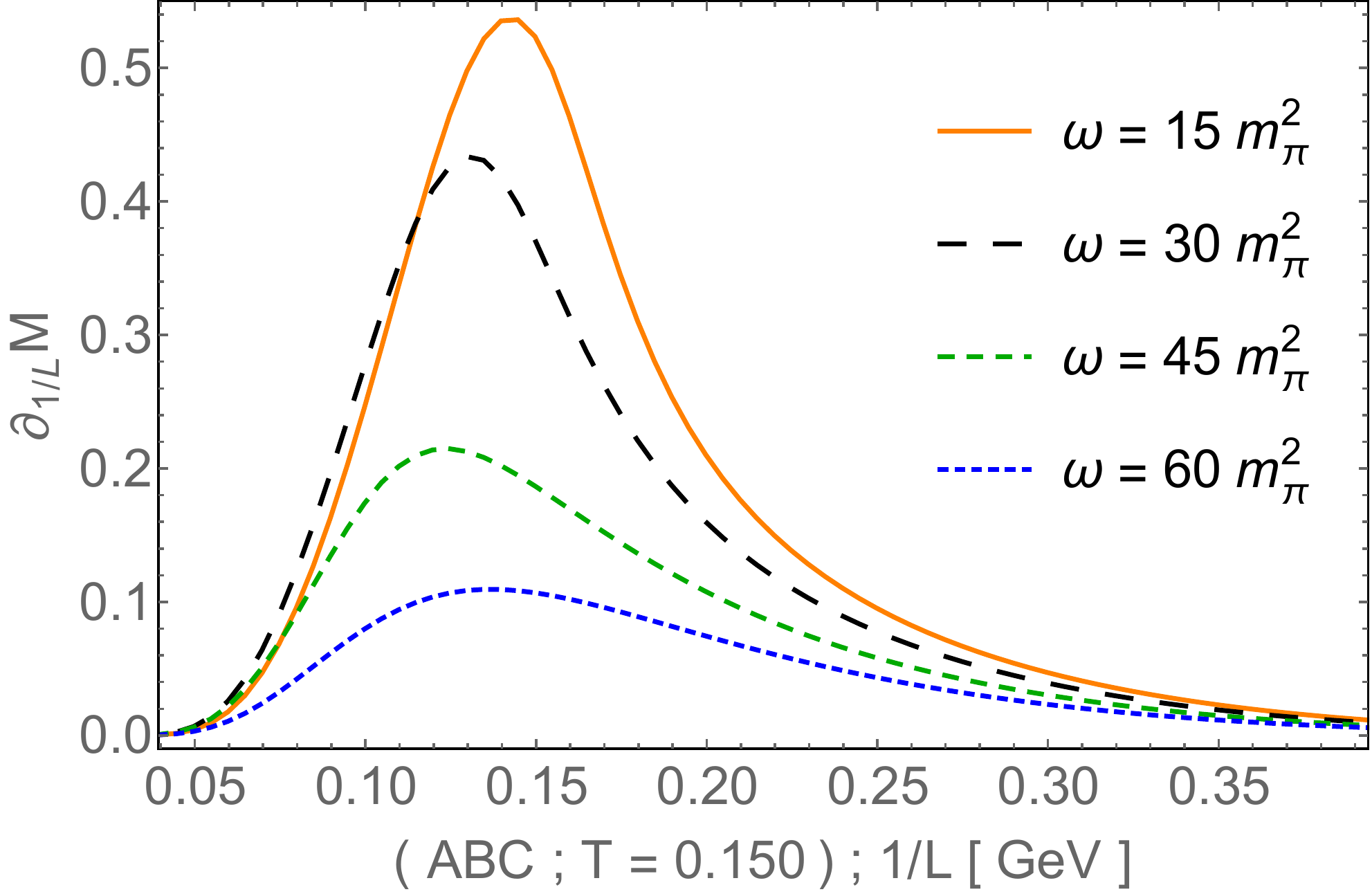}
\caption{Spatial susceptibility ($\partial M / \partial (L^{-1})$) as a function of the of inverse of length $ 1/L $ in ABC case, taking different values of the cyclotron frequency $ \omega $ and temperature $T$. Here we use the squared pion mass $ m_{\pi}^2 (\approx 0.018 \mathrm{GeV}^2)$ as scale for $ \omega $. }
\label{Susc3}
\end{figure}

\begin{figure}
\centering
\includegraphics[width=0.45\columnwidth]{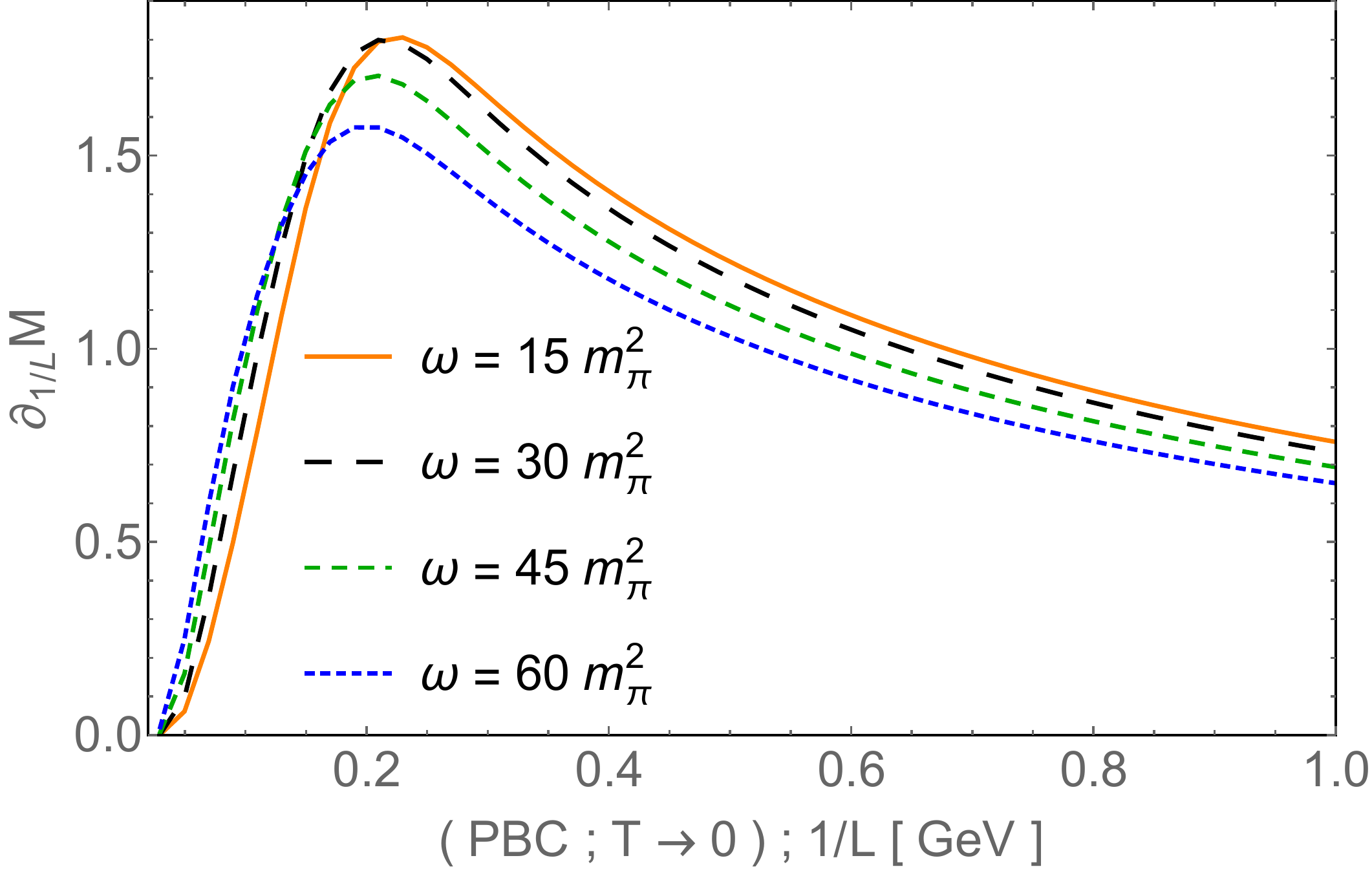}
\includegraphics[width=0.45\columnwidth]{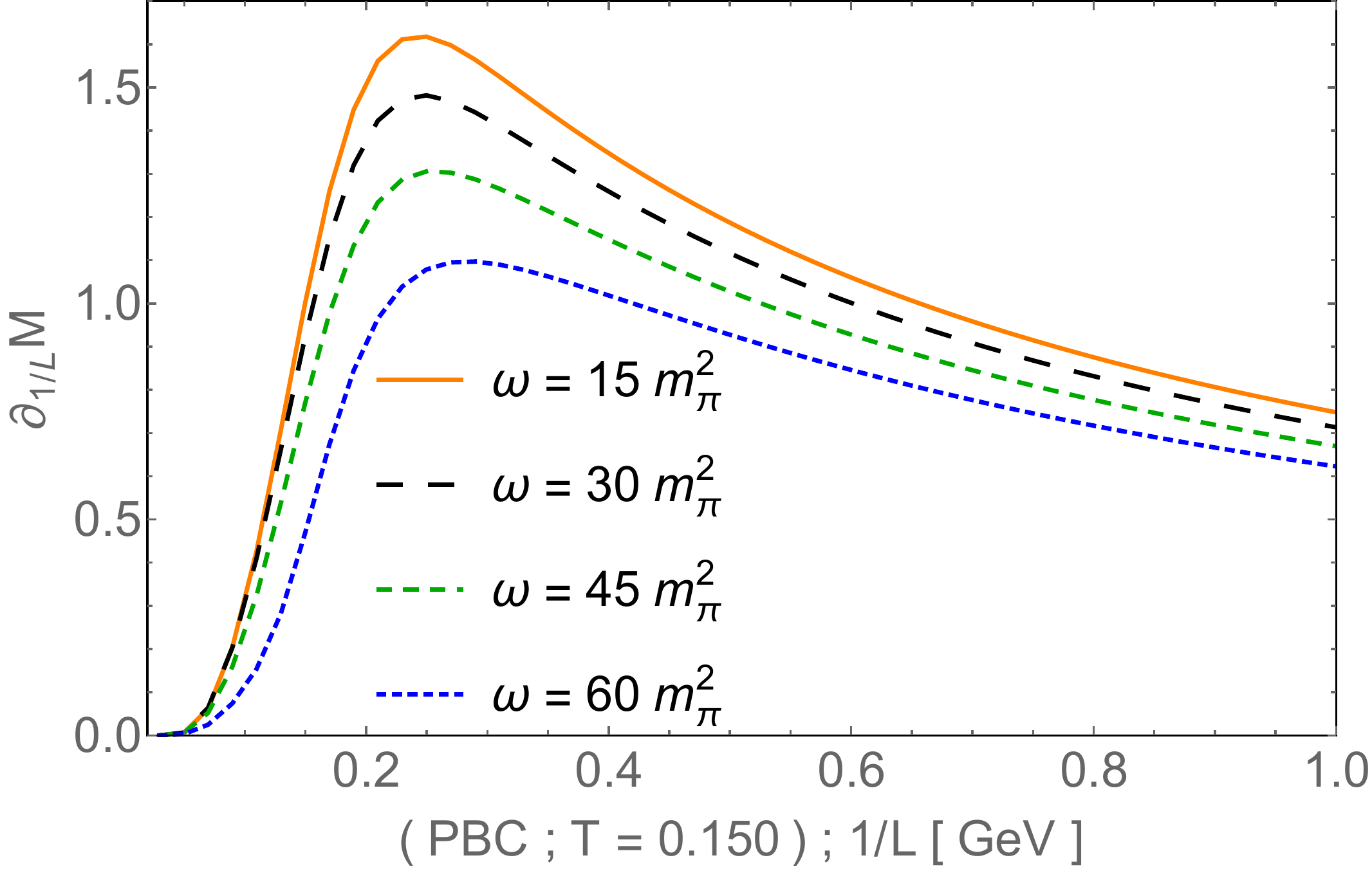}
\caption{The same as in Fig.~\ref{Susc3}, but in PBC case. }
\label{Susc4}
\end{figure}

For a more detailed assessment of the results presented above, in Figs.~\ref{Susc1}-\ref{Susc4} are plotted the thermal and spatial susceptibilities, defined as $\partial M / \partial T$ and $\partial M / \partial (L^{-1})$ respectively, taking different magnetic field strengths, in ABC and PBC cases. 

We look first at the ABC context (Figs.~\ref{Susc1} and \ref{Susc3}). The prominent peak of the plots designates the occurrence of the crossover transition. In this sense, the peak location of the thermal (spatial) susceptibility indicates the pseudo-critical temperature $T_c$  (inverse of pseudo-critical length $L_c^{-1}$). It can be seen from the curves associated to the limits of the range of magnetic field strength considered, that the increase of $\omega$ drops the height of the peaks as well as the values of $T_c$ and $L_c^{-1}$.  This is another way of regarding the manifestation of the IMC effect.  Additionally, the the combined size-magnetic effects in the ABC scenario pushes down the peaks of the thermal and spatial susceptibilities, contributing to the restoration of the symmetric phase.  

Under the circumstances of PBC (Figs.~\ref{Susc2} and \ref{Susc4}) the peak in the thermal susceptibility becomes sharper and moves to higher temperatures with the drop of $L$; but the increase of the magnetic field strength causes the opposite outcome. It means that while the enhancement of finite size effects engenders bigger $T_c$ (i.e. stimulation of the broken phase), the intensification of IMC produces smaller $T_c$ (in other words: the weakening of symmetry breaking).  The behavior of the spatial susceptibility complements this analysis: higher temperatures magnetic field strength mitigates the peak, but the decrease of the size does not generate a vanishing $\partial M / \partial (L^{-1})$. 
This might be interpreted as the absence of a critical value of the size in which the symmetry is restored. 

Hence, the main message of is work is that the chiral crossover transition, in the scenario of the NJL model with a magnetic dependent coupling constant used to reproduce the IMC, depends strongly on the combined thermo-size-magnetic effects, in particular on the boundary conditions adopted.

\section{Concluding remarks}

We have focused attention here on how the combined thermo-size-magnetic effects affect the the properties of the quarkionic matter, in the context of the two-flavor NJL model in the presence of a magnetic dependent coupling constant parametrized in consonance with the IMC predicted in LQCD. To this end, we have made use of the mean-field approximation, the Schwinger proper time method and a geneneralized Matsubara prescription. Our findings suggest that the phase diagram is strongly affected by the combined effects of the mentioned variables and, most interestingly, by the periodicity of the boundary conditions. 

The concomitance of antipediodic boundaries and IMC effects causes the inhibition of the dynamical breaking of chiral symmetry, with the system acquiring smaller  values of the constituent quark mass $M$, the pseudo-critical temperature $ T_c $, and the inverse of pseudo-critical length $L_c^{-1}$. But in contrast with the ABC case where thermo-size-magnetic effects attenuate and disfavor the broken phase, the PBC scenario is characterized by a concurrence between magnetic and finite-size effects: while the former lowers $M$  and $L_c^{-1}$, the latter acts oppositely, raising $M$ and $ T_c $, with net result depending on the balance of these competing conditions.

Finally, Let us mention some remarks on this approach. The outcomes obtained in this study obviously depend on the regularization procedure and parametrization choice. A distinct set of input parameters might modify the value of constituent quark mass and ranges of $(T,L, \omega)$ where thermo-size-magnetic effects becomes relevant. Notwithstanding, the findings reported in precedent sections  provide a better understanding on how the finite-volume and magnetic effects influences the strongly interacting matter produced in environments like heavy-ion collisions or in lattice simulations. 

\acknowledgments

The authors would like to thank the Brazilian funding agencies for their 
financial support: CNPq (LMA: contracts 309950/2020-1, 400215/2022-5), 
FAPESB (LMA: contract INT0007/2016) and the INCT-FNA.

%
%

%


\appendix

\section*{Appendix: Chiral condensate under MFIR regularization}

In this Appendix we derive the expressions (\ref{regabc}) and/ \ref{regpbc}) for the condensates defined in (\ref{phi3magABC}) and (\ref{phi3magPBC}) using the proper time method taking the so-called magnetic field independent regularization  (MFIR) procedure, where the finite magnetic contribution is disentangled from the non-magnetic infinite one and only the latter is regularized~\cite{Avancini:2019wed}. 
First, from the definition of Jacobi theta functions $\theta_{2}$ and $\theta_{3}$ it is possible to write them as
\begin{eqnarray}
\theta_{2}\left[\frac{2\pi \mu S}{\beta} ; \exp\left(-4\pi^2 S / \beta^{2}\right)\right] &=& \frac{\sqrt{\pi}\beta}{2\pi S^{1/2}}\exp\left(-\mu^{2}S\right) \nonumber \\
&\times&\left[1+2 \sum_{n_{\tau} = 1}^{+\infty}\exp\left(-\beta^{2}n^{2}_{\tau}/4S\right)(-1)^{n_{\tau}} \cosh{\left(n_{\tau}\beta \mu\right)}\right]
\end{eqnarray}
and
\begin{eqnarray}
\theta_{3}\left[0 ; \exp\left(-4\pi^2 S / L_{z}^{2}\right)\right] &=& \frac{\sqrt{\pi}L_{z}}{2\pi S^{1/2}} \left[1+2 \sum_{n_{z} = 1}^{+\infty} (-1)^{n_{z}} \exp\left(-\ L_{z}^{2}n^{2}_{z}/4S\right)\right]
\end{eqnarray}
As a consequence, after some manipulations the quark condensates for ABC case becomes
\begin{eqnarray}
\phi_f(\omega,T,L_{z},\mu) &=& \frac{2N_{c}M  \omega_f }{\pi} \frac{\pi}{4\pi^2} \sum_{\ell = 0}^{+\infty}\sum_{s=\pm1}^{} \int_{0}^{\infty} \frac{dS}{S} \left\{1+\left[2 \sum_{n_{z} = 1}^{+\infty} (-1)^{n_{z}} \exp\left(-L_{z}^{2} n^{2}_{z}/4S\right)\right]\right.\nonumber \\
&+&\left[2 \sum_{n_{\tau} = 1}^{+\infty} (-1)^{n_{\tau}} \exp\left(-\beta^{2}n^{2}_{\tau}/4S\right) \cosh{\left(n_{\tau}\beta \mu\right)}\right] \nonumber \\
&+&\left.\left[2^2 \sum_{n_{\tau,n_z} = 1}^{+\infty}\exp\left[-(\beta^{2}n^{2}_{\tau}+L^{2}_{z}n^{2}_{z})/4S\right](-1)^{n_{\tau}+n_{z}} \cosh{\left(n_{\tau}\beta \mu\right)}\right]\right\} \nonumber \\
&\times&\exp\left\{-S\left[M^{2}+\omega_f (2\ell+1 - s)\right]\right\}.
\label{phi3magABC2}
\end{eqnarray}
Using the relation
\ben
 \int_{0}^{\infty} {dS}\,{S^{\nu -1}} \, \exp\left[-\left(\frac{\alpha}{S}+\gamma S\right)\right] = 2 \left(\frac{\alpha}{\gamma}\right)^{\nu/2} \, K_{\nu}\left(2 \sqrt{\alpha \gamma}\right), 
\een
Eq.~(\ref{phi3magABC2}) takes the form
\begin{eqnarray}
\phi_f(\omega,T,L_{z},\mu) &=& \frac{2N_{c}M  }{\pi} \frac{\pi}{4\pi^2} \left\{ \omega_{f} \sum_{\ell = 0}^{+\infty}\sum_{s=\pm1}^{} \int_{0}^{\infty} \frac{dS}{S} \exp\left\{-S\left[M^{2}+\omega_f (2\ell+1 - s)\right]\right\} \right. \nonumber \\
&+& \omega_{f} \sum_{\ell = 0}^{+\infty}\sum_{s=\pm1}^{} \left[\left[2^{2} \sum_{n_{z} = 1}^{+\infty} (-1)^{n_{z}} K_{0}\left(L_{z} n_{z}\sqrt{M^{2}+\omega_{f}(2\ell+1-s)}\right) \right] \right. \nonumber \\
&+&\left[2^{2} \sum_{n_{\tau} = 1}^{+\infty} (-1)^{n_{\tau}} \cosh{(n_{\tau}\beta\mu)}K_{0}\left(\beta \,n_{\tau}\sqrt{M^{2}+\omega_{f}(2\ell+1-s)}\right) \right] \nonumber \\
&+&\left[2^{3} \sum_{n_{\tau},n_{z} = 1}^{+\infty} (-1)^{n_{\tau}+n_{z}} \cosh{(n_{\tau}\beta\mu)} \right.\nonumber \\
&\times&\left.\left.\left.K_{0}\left(\sqrt{\left(\beta^2 n_{\tau}^2 + L_{z}^{2} n_{z}^{2}\right)\left(M^{2}+\omega_{f}(2\ell+1-s)\right)}\right)\right]\right]\right\}.
\end{eqnarray}
Then, after performing the summation over the spin and Landau levels in the term in second line of equation above, we obtain (already including the cutoff in the integral)
\begin{eqnarray}
\phi_f(\omega,T,L_{z},\mu) &=& \frac{2N_{c}M  }{\pi} \frac{\pi}{4\pi^2} \left\{  \int_{1/\Lambda^{2}}^{\infty} \frac{dS}{S^{2}} \exp\left(-S \,M^{2}\right) \right. \nonumber \\
&+& \omega_{f}\left[ 2 \, \zeta^{\prime}(0,x_{f})+(1-2x_{f})\ln{x_{f}}+2x_{f}  \right] \nonumber \\
&+& \omega_{f} \sum_{\ell = 0}^{+\infty}\sum_{s=\pm1}^{} \left[\left[2^{2} \sum_{n_{z} = 1}^{+\infty} (-1)^{n_{z}} K_{0}\left(L_{z} n_{z}\sqrt{M^{2}+\omega_{f}(2\ell+1-s)}\right) \right] \right. \nonumber \\
&+&\left[2^{2} \sum_{n_{\tau} = 1}^{+\infty} (-1)^{n_{\tau}} \cosh{(n_{\tau}\beta\mu)}K_{0}\left(\beta \,n_{\tau}\sqrt{M^{2}+\omega_{f}(2\ell+1-s)}\right) \right] \nonumber \\
&+&\left[2^{3} \sum_{n_{\tau},n_{z} = 1}^{+\infty} (-1)^{n_{\tau}+n_{z}} \cosh{(n_{\tau}\beta\mu)} \right.\nonumber \\
&\times & \left.\left.\left.K_{0}\left(\sqrt{\left(\beta^2 n_{\tau}^2 + L_{z}^{2} n_{z}^{2}\right)\left(M^{2}+\omega_{f}(2\ell+1-s)\right)}\right)\right]\right]\right\},
\end{eqnarray}
where $x_{f} \equiv M^{2} / 2\omega_{f}$ and $\zeta^{\prime}(0,x_{f})$ is the derivative of the Hurwitz zeta function with respect to the first argument.

Thus, proceeding as before but for the PBC case, we get
\begin{eqnarray}
\phi_f(\omega,T,L_{z},\mu) &=& \frac{2N_{c}M  }{\pi} \frac{\pi}{4\pi^2} \left\{  \int_{1/\Lambda^{2}}^{\infty} \frac{dS}{S^{2}} \exp\left(-S \,M^{2}\right) \right. \nonumber \\
&+& \omega_{f}\left[2 \, \zeta^{\prime}(0,x_{f})+(1-2x_{f})\ln{x_{f}}+2x_{f} \frac{}{}\right] \nonumber \\
&+& \omega_{f} \sum_{\ell = 0}^{+\infty}\sum_{s=\pm1}^{} \left[\left[2^{2} \sum_{n_{z} = 1}^{+\infty}  K_{0}\left(L_{z} n_{z}\sqrt{M^{2}+\omega_{f}(2\ell+1-s)}\right) \right] \right. \nonumber \\
&+&\left[2^{2} \sum_{n_{\tau} = 1}^{+\infty} (-1)^{n_{\tau}} \cosh{(n_{\tau}\beta\mu)}K_{0}\left(\beta \,n_{\tau}\sqrt{M^{2}+\omega_{f}(2\ell+1-s)}\right) \right] \nonumber \\
&+&\left[2^{3} \sum_{n_{\tau},n_{z} = 1}^{+\infty} (-1)^{n_{\tau}} \cosh{(n_{\tau}\beta\mu)} \right.\nonumber \\
&\times&\left.\left.\left.K_{0}\left(\sqrt{\left(\beta^2 n_{\tau}^2 + L_{z}^{2} n_{z}^{2}\right)\left(M^{2}+\omega_{f}(2\ell+1-s)\right)}\right)\right]\right]\right\}.
\end{eqnarray}

\bibliographystyle{plain}

\begin{thebibliography}{99}


\bibitem{rev-qgp}      P. Braun-Munzinger, V. Koch,
                       T. Schafer, and J. Stachel,
                       Phys. Rep. {\bf 621}, 76 (2016).


\bibitem{Prino:2016cni} 
  F.~Prino and R.~Rapp,
  J.\ Phys.\ G {\bf 43}, no. 9, 093002 (2016).



\bibitem{Pasechnik:2016wkt} 
   R.~Pasechnik and M.~Sumbera,
  Universe {\bf 3}, no. 1, 7 (2017).


\bibitem{Kharzeev} D.E Kharzeev, L.D. Mclerran and H.J. Warringa, ArXiv:0711.0950 [hep-ph], 2007.
  
\bibitem{Skokov:2009qp} V.~Skokov, A.~Y.~Illarionov and V.~Toneev, Int.\ J.\ Mod.\ Phys.\ A {\bf 24}, 5925 (2009).  

\bibitem{Chernodub:2010qx} M.~N.~Chernodub,
  Phys.\ Rev.\ D {\bf 82}, 085011 (2010).


\bibitem{Ayala1} A. Ayala, M. Loewe, J.C. Rojas and C. Villavicencio, Phys. Rev. D \textbf{86}, 076006 (2012).


 \bibitem{Tobias} M. Ferreira, P. Costa, O. Lourenco, T. Frederico, and C. Providencia,  Phys. Rev. D {\bf 89}, 116011 (2014).
  
\bibitem{Heber} A. Haber, F. Preis and A. Schmitt, Phys. Rev. D \textbf{90}, 125036 (2014).


\bibitem{Mamo:2015dea}  K.~A.~Mamo,
  JHEP {\bf 1505}, 121 (2015)

\bibitem{MAO} S. Mao, Phys. Lett. B {\bf 758}, 195 (2016).

\bibitem{Ayala2} A. Ayala, P. Mercado and C. Villavicencio, Phys. Rev. C \textbf{95}, 014904 (2017).



\bibitem{Zhang:2016qrl}
R.~Zhang, W.~j.~Fu and Y.~x.~Liu,
Eur. Phys. J. C \textbf{76}, no.6, 307 (2016)
doi:10.1140/epjc/s10052-016-4123-8

\bibitem{Farias:2016gmy}
R.~L.~S.~Farias, V.~S.~Timoteo, S.~S.~Avancini, M.~B.~Pinto and G.~Krein,
Eur. Phys. J. A \textbf{53} (2017) no.5, 101
doi:10.1140/epja/i2017-12320-8
[arXiv:1603.03847 [hep-ph]].


\bibitem{Pagura} V. P. Pagura, D. Gomez Dumm, S. Noguera, and N. N. Scoccola
Phys. Rev. D {\bf 95}, 034013 (2017).



\bibitem{Magdy} N. Magdy, M. Csanad and R. A. Lacey, J. Phys. G: Nucl. Part. Phys. {\bf 44}, 025101 (2017).


\bibitem{Ayala0} A. Ayala, C. A. Dominguez, S. Hernandez-Ortiz, L. A. Hernandez, M. Loewe, D. M. Paret, and R. Zamora, Phys. Rev. D {\bf 98}, 031501(R) (2018).



\bibitem{Wang:2017vtn}
Z.~Wang and P.~Zhuang,
Phys. Rev. D \textbf{97}, no.3, 034026 (2018)
doi:10.1103/PhysRevD.97.034026

\bibitem{Martinez:2018snm}
A.~Mart\'\i{}nez and A.~Raya,
Nucl. Phys. B \textbf{934}, 317-329 (2018)
doi:10.1016/j.nuclphysb.2018.07.008
[arXiv:1804.03183 [hep-th]].


\bibitem{Mao:2018dqe}
S.~Mao,
Phys. Rev. D \textbf{99}, no.5, 056005 (2019)
doi:10.1103/PhysRevD.99.056005

\bibitem{Avancini:2018svs}
S.~S.~Avancini, R.~L.~S.~Farias and W.~R.~Tavares,
Phys. Rev. D \textbf{99}, no.5, 056009 (2019)
doi:10.1103/PhysRevD.99.056009

\bibitem{Avancini:2019wed}
S.~S.~Avancini, R.~L.~S.~Farias, N.~N.~Scoccola and W.~R.~Tavares,
Phys. Rev. D \textbf{99} (2019) no.11, 116002
doi:10.1103/PhysRevD.99.116002
[arXiv:1904.02730 [hep-ph]].

\bibitem{Abreu:2019czp} 
  L.~M.~Abreu, E.~B.~S.~Correa, C.~A.~Linhares and A.~P.~C.~Malbouisson,
  Phys.\ Rev.\ D {\bf 99}, no. 7, 076001 (2019)

\bibitem{Ghosh:2021dlo}
S.~Ghosh, N.~Chaudhuri, P.~Roy and S.~Sarkar,
Phys. Rev. D \textbf{103}, 116008 (2021)
doi:10.1103/PhysRevD.103.116008


\bibitem{Abreu:2021btt}
L.~M.~Abreu, E.~B.~S.~Corr\^ea and E.~S.~Nery,
Phys. Rev. D \textbf{105} (2022) no.5, 056010
doi:10.1103/PhysRevD.105.056010
[arXiv:2109.09593 [hep-ph]].



%
%

%
%
%


%
%
%
%
%


%
%
%
%


%
%
%
%
%
%
%
%
%
%
%
%
%
%
%
%



%


\bibitem{Bali:2012zg}
G.~S.~Bali, F.~Bruckmann, G.~Endrodi, Z.~Fodor, S.~D.~Katz and A.~Schafer,
Phys. Rev. D \textbf{86} (2012), 071502
doi:10.1103/PhysRevD.86.071502
[arXiv:1206.4205 [hep-lat]].

\bibitem{Bali:2011qj}
G.~S.~Bali, F.~Bruckmann, G.~Endrodi, Z.~Fodor, S.~D.~Katz, S.~Krieg, A.~Schafer and K.~K.~Szabo,
JHEP \textbf{02} (2012), 044
doi:10.1007/JHEP02(2012)044
[arXiv:1111.4956 [hep-lat]].

\bibitem{Farias:2014eca}
R.~L.~S.~Farias, K.~P.~Gomes, G.~I.~Krein and M.~B.~Pinto,
Phys. Rev. C \textbf{90} (2014) no.2, 025203
doi:10.1103/PhysRevC.90.025203
[arXiv:1404.3931 [hep-ph]].

\bibitem{Ferreira:2014kpa}
M.~Ferreira, P.~Costa, O.~Louren\c{c}o, T.~Frederico and C.~Provid\^encia,
Phys. Rev. D \textbf{89} (2014) no.11, 116011
doi:10.1103/PhysRevD.89.116011
[arXiv:1404.5577 [hep-ph]].

\bibitem{Ferreira:2017wtx}
M.~Ferreira, P.~Costa and C.~Provid\^encia,
Phys. Rev. D \textbf{97} (2018) no.1, 014014
doi:10.1103/PhysRevD.97.014014
[arXiv:1712.08378 [hep-ph]].

\bibitem{Ahmad:2016iez}
A.~Ahmad and A.~Raya,
J. Phys. G \textbf{43} (2016) no.6, 065002
doi:10.1088/0954-3899/43/6/065002
[arXiv:1602.06448 [hep-ph]].

\bibitem{Andersen:2021lnk}
J.~O.~Andersen,
Eur. Phys. J. A \textbf{57} (2021) no.6, 189
doi:10.1140/epja/s10050-021-00491-y
[arXiv:2102.13165 [hep-ph]].
%




\bibitem{Bass:1998qm} 
  S.~A.~Bass {\it et al.},
  Prog.\ Part.\ Nucl.\ Phys.\  {\bf 42}, 313 (1999). 
  doi:10.1016/S0146-6410(99)00086-1
   
\bibitem{Palhares:2009tf} 
  L.~F.~Palhares, E.~S.~Fraga and T.~Kodama,
  J.\ Phys.\ G {\bf 38}, 085101 (2011). 
  doi:10.1088/0954-3899/38/8/085101
  
  
\bibitem{Graef:2012sh} 
  G.~Graf, M.~Bleicher and Q.~Li,
  Phys.\ Rev.\ C {\bf 85}, 044901 (2012).
  doi:10.1103/PhysRevC.85.044901
  
\bibitem{Shi:2018swj} 
  C.~Shi, W.~Jia, A.~Sun, L.~Zhang and H.~Zong,
  Chin.\ Phys.\ C {\bf 42}, no. 2, 023101 (2018).
  doi:10.1088/1674-1137/42/2/023101
  
\bibitem{Luecker:2009bs} 
  J.~Luecker, C.~S.~Fischer and R.~Williams,
  Phys.\ Rev.\ D {\bf 81}, 094005 (2010).
  doi:10.1103/PhysRevD.81.094005
  
\bibitem{Li:2017zny} 
  B.~L.~Li, Z.~F.~Cui, B.~W.~Zhou, S.~An, L.~P.~Zhang and H.~S.~Zong,
  Nucl.\ Phys.\ B {\bf 938}, 298 (2019).
  doi:10.1016/j.nuclphysb.2018.11.015

 
\bibitem{Braun:2004yk} 
  J.~Braun, B.~Klein and H.-J.~Pirner,
  Phys.\ Rev.\ D {\bf 71}, 014032 (2005)
  doi:10.1103/PhysRevD.71.014032
  
\bibitem{Braun:2005fj} 
  J.~Braun, B.~Klein, H.-J.~Pirner and A.~H.~Rezaeian,
  Phys.\ Rev.\ D {\bf 73}, 074010 (2006)
  doi:10.1103/PhysRevD.73.074010
 
 \bibitem{Ferrer:1999gs} 
  E.~J.~Ferrer, V.~P.~Gusynin and V.~de la Incera,
  Phys.\ Lett.\ B {\bf 455}, 217 (1999)
  [hep-ph/9901446].
  
  
 \bibitem{Abreu:2006}
  L.~M.~Abreu, M.~Gomes and A.~J.~da~Silva,
  {\it  Phys. Lett. B } {\bf 642}, 551 (2006).

\bibitem{Ebert0} D. Ebert, K. G. Klimenko, A. V. Tyukov and V. Ch. Zhukovsky, {\it Phys. Rev. D} {\bf 78}, 045008 (2008).

\bibitem{Abreu:2009zz}
  L.~M.~Abreu, A.~P.~C.~Malbouisson, J.~M.~C.~Malbouisson and A.~E.~Santana,
   {\it Nucl.\ Phys.\ B } {\bf 819}, 127 (2009).
  

\bibitem{Abreu:2011rj}
  L.~M.~Abreu, A.~P.~C.~Malbouisson and J.~M.~C.~Malbouisson,
  {\it  Phys.\ Rev.\ D } {\bf 83}, 025001 (2011).

\bibitem{Fraga}  E. S. Fraga and L. F.  Palhares, Phys. Rev. D {\bf 86}, 016008 (2012). 

\bibitem{Abreu3} L. M. Abreu, C. A. Linhares, A. P. C. Malbouisson, J. M. C. Malbouisson,  {\it Phys. Rev. D} {\bf  88}, 107701 (2013).


\bibitem{Bhattacharyya:2012rp} 
  A.~Bhattacharyya, P.~Deb, S.~K.~Ghosh, R.~Ray and S.~Sur,
  Phys.\ Rev.\ D {\bf 87}, no. 5, 054009 (2013).
 
 \bibitem{Abreu6} L. M. Abreu, A. P. C. Malbouisson, J. M. C. Malbouisson, E. S. Nery and R. Rodrigues da Silva, Nucl. Phys. B \textbf{881}, 327-342 (2014).

 
\bibitem{Bhattacharyya:2014uxa} A.~Bhattacharyya, R.~Ray and S.~Sur,
  Phys.\ Rev.\ D {\bf 91}, 051501(R) (2015).



\bibitem{Bhattacharyya2} A. Bhattacharyya, R. Ray, S. Samanta and S. Sur, Phys. Rev. C {\bf 91}, 041901(R) (2015).



\bibitem{Ebert3} D. Ebert, T. G. Khunjua, K. G. Klimenko and V. C. Zhukovsky,  {\it Phys. Rev. D} {\bf 91}, 105024 (2015). 

\bibitem{Abreu4} L. M. Abreu, E. S. Nery and A. P. C. Malbouisson, Phys. Rev. D \textbf{91}, 087701 (2015).

\bibitem{Abreu5} L. M. Abreu and E. S. Nery, Int. J. Mod. Phys. A {\bf 31}, 1650128 (2016). 

\bibitem{Abreu7} L. M. Abreu, A. P. C. Malbouisson and E. S. Nery, Mod. Phys. Lett. A {\bf 31}, 1650121 (2016).

\bibitem{Bao1} S.S. Bao and H. Shen, Phys. Rev. C {\bf 93}, 025807 (2016).

%

%
%

 \bibitem{Damgaard:2008zs} 
  P.~H.~Damgaard and H.~Fukaya,
  JHEP {\bf 0901}, 052 (2009)
  doi:10.1088/1126-6708/2009/01/052

\bibitem{Kohyama:2016fif} 
  H.~Kohyama, D.~Kimura and T.~Inagaki,
  Nucl.\ Phys.\ B {\bf 906}, 524 (2016)


\bibitem{PhysRevC.96.055204} L. M. Abreu, and E. S. Nery, {\it  Phys.\ Rev.\ C }
 {\bf 96}, 055204 (2017).  

\bibitem{Samanta} S. Samanta, S. Ghosh and B. Mohanty, J. Phys. G: Nucl. Part. Phys. {\bf 45}, 075101 (2018).

\bibitem{Wu} X.H. Wu and H. Shen, Phys. Rev. C \textbf{96}, 025802 (2017).

%


\bibitem{Pan:2016ecs} 
  Z.~Pan, Z.~F.~Cui, C.~H.~Chang and H.~S.~Zong,
  Int.\ J.\ Mod.\ Phys.\ A {\bf 32}, no. 13, 1750067 (2017).
  doi:10.1142/S0217751X17500671
  
\bibitem{Wang:2018kgj} 
  Q.~Wang, Y.~Xiq and H.~Zong,
  Mod.\ Phys.\ Lett.\ A {\bf 33}, no. 39, 1850232 (2018).
  doi:10.1142/S0217732318502322


\bibitem{Shi} C. Shi, Y. Xia, W. Jia et al., Sci. China Phys. Mech. Astron.  {\bf 61} 082021 (2018). 


\bibitem{XiaYongHui:2019gci} 
  Y.~Xia, Q.~Wang, H.~Feng and H.~Zong,
  Chin.\ Phys.\ C {\bf 43}, no. 3, 034101 (2019).
   doi:10.1088/1674-1137/43/3/034101
     
\bibitem{Abreu:2019tnf} 
  L.~M.~Abreu and E.~S.~Nery,
  Eur.\ Phys.\ J.\ A {\bf 55}, no. 7, 108 (2019)
  doi:10.1140/epja/i2019-12793-3
	  

\bibitem{Abreu:2020uxc}
L.~M.~Abreu, E.~S.~Nery and E.~B.~S.~Corr\^ea,
Physica A \textbf{572}, 125885 (2021)
doi:10.1016/j.physa.2021.125885


 \bibitem{Das:2019crc} 
  A.~Das, D.~Kumar and H.~Mishra,
  Phys.\ Rev.\ D {\bf 100}, no. 9, 094030 (2019)
  doi:10.1103/PhysRevD.100.094030
 


\bibitem{Zhao:2018nqa}
Y.~P.~Zhao, P.~L.~Yin, Z.~H.~Yu and H.~S.~Zong,
Nucl. Phys. B \textbf{952}, 114919 (2020)
doi:10.1016/j.nuclphysb.2020.114919

\bibitem{Klein:2017shl} 
  B.~Klein,
  Phys.\ Rept.\  {\bf 707-708}, 1 (2017).
  doi:10.1016/j.physrep.2017.09.002


\bibitem{NJL} Y. Nambu and G. Jona-Lasinio, Phys. Rev. \textbf{122}, 345 (1961).

\bibitem{NJL1} Y. Nambu and G. Jona-Lasinio, Phys. Rev. \textbf{124}, 246 (1961).

\bibitem{Vogl} U. Vogl and W. Weise, Prog. Part. Nucl. Phys. \textbf{27}, 195 (1991).

\bibitem{Klevansky}  S. P. Klevansky, Rev. Mod. Phys. {\bf 64}, 649 (1992).

\bibitem{Hatsuda} T. Hatsuda and T. Kunihiro, Phys. Rep. \textbf{247}, 221 (1994).

\bibitem{Buballa} M. Buballa, Phys. Rep. \textbf{407}, 205 (2005).




\bibitem{livro} F.C. Khanna, A.P.C. Malbouisson, J.M.C. Malbouisson, and A.E.
Santana, {\it{Thermal Quantum Field Theory: Algebraic Aspects and Applications}}%
, World Scientific, Singapore (2009).

\bibitem{PR2014} F.C. Khanna, A.P.C. Malbouisson, J.M.C. Malbouisson, and A.E.
Santana, Phys. Rep. \textbf{539}, 135 (2014). 

\bibitem{Emerson} E. B. S. Corr\^ea, C. A. Linhares, A. P. C. Malbouisson, J. M. C. Malbouisson, and A. E. Santana, Eur. Phys. J. C, \textbf{77}, 261 (2017).


\bibitem{Schwinger} J. Schwinger, Phys. Rev. \textbf{82}, 664 (1951).

\bibitem{DeWitt1}  B. DeWitt, Dynamical Theory of Groups and Fields. Gordon and Breach, New York (1965).

\bibitem{DeWitt2} B. DeWitt, Phys. Rep. \textbf{19C}, 295 (1975). 

\bibitem{Ball} R. D. Ball, Phys. Rep. \textbf{182}, 1 (1989).   



\bibitem{Bellman} R. Bellman, {\it{A Brief Introduction to Theta Functions}}, Holt, Rinehart and Winston, Inc., New York (1961).

\bibitem{Mumford} D. Mumford,  {\it{Tata lectures on theta}}, Boston-Basel-Stuttgart: Birkhauser, vol. 1, (1983), vol. 2, (1984).









\bibitem{Miransky} V. A. Miransky and I. A. Shovkovy, Phys. Rev. D. \textbf{66}, 045006 (2002).


\bibitem{Cloet:2014rja}
I.~C.~Clo\"et, W.~Bentz and A.~W.~Thomas,
Phys. Rev. C \textbf{90} (2014), 045202
doi:10.1103/PhysRevC.90.045202
[arXiv:1405.5542 [nucl-th]].

\bibitem{Zhang2016MPLA} Zhang, J.-L., Shi, Y.-M., Xu, S.-S. et al., Mod. Phys. Let. A, \textbf{31}, 1650086 (2016). doi:10.1142/S0217732316500863

\bibitem{Hutauruk:2018zfk}
P.~T.~P.~Hutauruk, W.~Bentz, I.~C.~Clo\"et and A.~W.~Thomas,
Phys. Rev. C \textbf{97} (2018) no.5, 055210
doi:10.1103/PhysRevC.97.055210
[arXiv:1802.05511 [nucl-th]].


\bibitem{Ishikawa:1996jb}
K.~I.~Ishikawa, T.~Inagaki, K.~Fukazawa and K.~Yamamoto,
[arXiv:hep-th/9609018 [hep-th]].

%






\end{thebibliography}

\end{document}